\documentclass[iop,floatfix,numberedappendix]{emulateapj}
\pdfoutput=1

\usepackage[titletoc,title]{appendix}

\RequirePackage{color}
\usepackage{hyperref}

\usepackage{psfig}

\def\plotext{pdf}

\def\picdir{.}

\shorttitle{HSC Version 1}
\shortauthors{Whitmore et al.}

\begin{document}

\title{Version 1 of the Hubble Source Catalog}

\author{
Bradley C.\ Whitmore,\altaffilmark{1}
Sahar S.\ Allam,\altaffilmark{1,2}
Tam{\'a}s Budav{\'a}ri,\altaffilmark{3}
Stefano Casertano,\altaffilmark{1}
Ronald A. Downes,\altaffilmark{1}\break
Thomas Donaldson,\altaffilmark{1}
S.~Michael Fall,\altaffilmark{1}
Stephen H.\ Lubow,\altaffilmark{1}
Lee Quick,\altaffilmark{1}
Louis-Gregory Strolger,\altaffilmark{1}\break
Geoff Wallace,\altaffilmark{1}
Richard L.\ White\altaffilmark{1}
 }
\altaffiltext{1}{Space Telescope Science Institute (STScI), 3700 San Martin Drive, Baltimore, MD 21218.}
\altaffiltext{2}{Fermi National Accelerator Laboratory, P.O.\ Box 500, Batavia, IL 60510.}
\altaffiltext{3}{Center for Astrophysical Sciences, Department of Physics and Astronomy, Johns Hopkins University, 3400 North Charles Street, Baltimore, MD 21218.}

\begin{abstract}
The Hubble Source Catalog is designed to help optimize science from
the \textit{Hubble Space Telescope} by combining the tens of thousands
of visit-based source lists in the Hubble Legacy Archive into a
single master catalog. Version 1 of the Hubble Source Catalog
includes WFPC2, ACS/WFC, WFC3/UVIS, and WFC3/IR photometric data
generated using SExtractor software to produce the individual source
lists.  The catalog includes roughly 80~million
detections of 30~million objects involving 112 different detector/filter
combinations, and about 160~thousand \textit{HST} exposures.  Source lists
from Data Release~8 of the Hubble Legacy Archive are matched using
an algorithm developed by Budav{\'a}ri \& Lubow (2012).  The mean
photometric accuracy for the catalog as a whole is better than 0.10~mag, 
with relative accuracy as good as 0.02~mag in certain circumstances
(e.g., bright isolated stars). The relative astrometric residuals
are typically within 10~mas, with a value for the mode (i.e., most
common value) of 2.3~mas. The absolute astrometric accuracy is
better than $\sim$0.1~arcsec for most sources, but can be much larger for
a fraction of fields that could not be matched to the PanSTARRS,
SDSS, or 2MASS reference systems. In this paper we describe the
database design with emphasis on those aspects that enable the users
to fully exploit the catalog while avoiding common misunderstandings
and potential pitfalls.  We provide usage examples to illustrate
some of the science capabilities and data quality characteristics,
and briefly discuss plans for future improvements to the Hubble
Source Catalog.
\end{abstract}


\maketitle

\section{Introduction}\label{sec:intro}

The \textit{Hubble Space Telescope} (\textit{HST}) has been in orbit for over
25 years. In that time it has observed with a dozen different
instruments, hundreds of observing modes, and roughly a thousand
different filters and gratings. Selected, effectively pencil-beam
observations have been taken of only a small fraction of the total
sky, with a range of exposure times from less than a second (e.g.,
searches for faint companion planets around very bright stars), to
week-long observations of ``blank'' parts of the sky to observe
galaxies at the edge of the universe. This diversity reflects both
the promise and the challenge of the Hubble Source Catalog (HSC).

In recent times, computer-based catalogs of astronomical objects
have proven to be of great benefit to astronomers (e.g., the Sloan
Digital Sky Survey; SDSS, e.g., Ahn et~al.\ 2014).  By querying such 
databases, astronomers are able to carry out research that would otherwise 
be very time-consuming or completely impractical.  Taking a page from this
book, the HSC is designed to eventually include the majority of all the objects
ever observed by \textit{HST} into a single master catalog. Presently, and 
primarily due to limitations in the older WFPC2 and ACS source lists which 
do not go as deep as possible (as discussed in \S5), Version~1 contains roughly 
20\% of the objects it will eventually include. For example, improvements made 
for Version~2 will increase the size of the HSC by about a factor of three.

\begin{figure}[b]
\centering
\includegraphics[type=\plotext,ext=.\plotext,read=.\plotext,width=\linewidth]{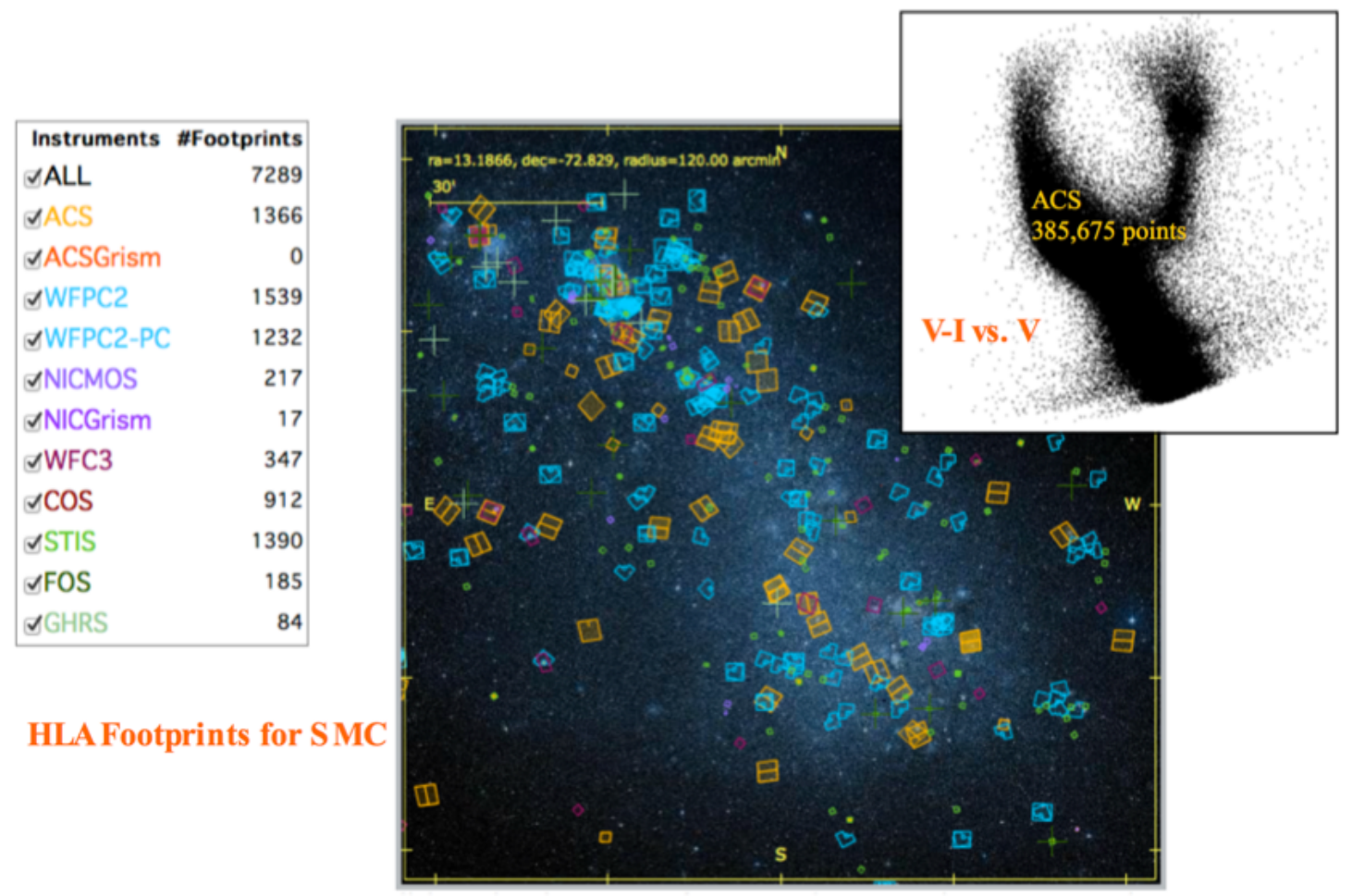}
\caption{ HLA footprints for a search of the SMC using a radius of 2~degrees. A color-magnitude 
diagram using the ACS-F606W (V) and ACS-F814W (I) filters, containing 385,675 data points, and 
created by the HSC in less than 2~minutes, is shown in the upper right.
\label{fig-SMC}}
\end{figure}

Repeat observations are common for \textit{HST}, with 500,000 objects having 
more than 50 separate observations and 8~million objects observed in more than 
10 separate observations. This provides a rich database for variability studies. 
Regions of the sky with thousands, or even tens of thousands of separate 
observations (e.g., the Magellanic Clouds---see Figure~\ref{fig-SMC}, the 
Virgo cluster, the Orion Nebula, M31, etc.) can be evaluated in minutes.

The basic scheduling unit for an \textit{HST} observation is a ``visit,'' typically lasting 
between a single orbit (96~min) and six or seven orbits.  A visit is also a natural unit 
for the production of data products from the telescope. For this reason, the Hubble Legacy
Archive (HLA, see Jenkner et~al.\ 2006; Whitmore et~al.\ 2008) combines data together 
in visit-based images and produces source lists for each of these combined images.

\begin{figure*}[t]
\centering
\includegraphics[type=\plotext,ext=.\plotext,read=.\plotext,width=0.7\linewidth]{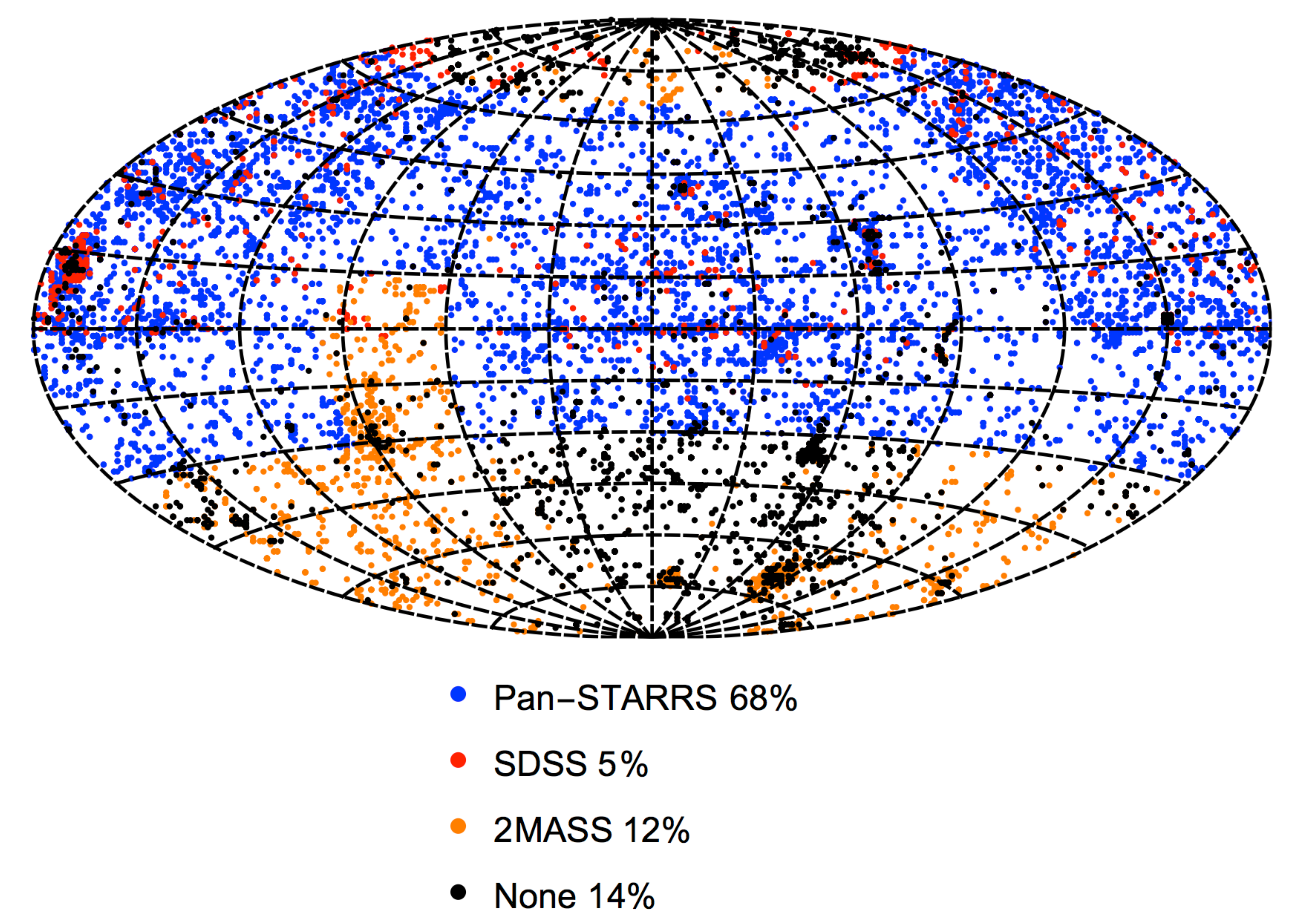}
\caption{Piecemeal sky coverage showing HLA images used to build the HSC. Color coding 
shows where PanSTARRS (pre-release version---PV1---e.g., see Stubbs et~al.\ 2010), SDSS 
(Ahn et~al.\ 2014), and 2MASS (Skrutskie et~al.\ 2006) are used to provide the astrometric 
backbone for the HSC.
\label{fig-coverage}}
\end{figure*}

In general, an astronomer is not interested in visits, but would
like to retrieve all the relevant information for a target observed
by \textit{Hubble}. That is the primary driver behind the production
of the HSC: to combine the tens of thousands of visit-based HLA
source lists into a single master catalog.

The HSC has been available as a Beta (test) version since 2012.
Special purpose techniques were developed to handle the challenges
of building the HSC.  The pipeline, the astrometric and cross-matching
algorithms, and the properties of the Beta version of the catalog
are described in Budav{\'a}ri \& Lubow (2012).  In the current
paper, we describe Version~1 of the HSC.  We provide a brief update
on the catalog generation methods and the catalog properties since
the original Beta release.

Many astronomical catalogs are produced by telescopes that conduct
systematic surveys. The catalog is a key objective of the survey
and the observations cover a regular geometric pattern in the sky
with uniform properties, such as exposure time and filter set.  \textit{The 
HSC is a very different type of catalog}, as illustrated by
Figure~\ref{fig-SMC}. Due to the diversity of  \textit{Hubble} observations,
and accentuated by the fact that the HSC is still in an active
developmental stage, the catalog can be very non-uniform, with a
patchwork nature in certain regions.  This irregularity requires
care when developing search criteria. Nevertheless, the HSC is a
powerful tool for research with \textit{Hubble} data, even with its
limitations, and will be an important reference for future telescopes,
such as the \textit{James Webb Space Telescope} (\textit{JWST}), and
for survey programs such as PanSTARRS (Panoramic Survey Telescope 
and Rapid Response System) and LSST (Large Synoptic Survey Telescope).

The HSC is not designed to be all things to all people. The HLA source lists 
are meant for general-usage rather than being tuned for a particular data set 
or science requirement. In most cases, a higher science return is possible using 
a specialized catalog designed for specific needs (e.g., crowded field photometry) 
or specific science projects (e.g., the Ultra Deep Field). In particular, the High 
Level Science Products (HLSP; \url{http://archive.stsci.edu/hlsp/}) produced 
by individual science teams are generally of higher quality than the HSC. 

The paper is organized as follows. In Section~2 we describe the data used in 
building Version~1 of the HSC, while in Section~3 we describe the pipeline 
used to construct the catalog. In Section~4 we examine the photometric and 
astrometric quality of the HSC.  Sections~5 includes advice on avoiding 
common misunderstandings and potential pitfalls. Section~6 gives a brief 
summary and describes future plans. The appendices provide  comparisons with 
studies based on use cases, describe tools that can be used to query the HSC, 
and provide pointers to other relevant information.

\newpage\section{The Data }\label{sec:data}
\subsection{Instruments and Filters}
Version 1 of the HSC includes HLA source lists from the three cameras
responsible for the majority of images taken by \textit{Hubble}, namely
the Wide Field Planetary Camera 2 (WFPC2), the Wide Field Camera
of the Advanced Camera for Surveys (ACS/WFC), and both the
ultraviolet/visible and infrared channels of the Wide Field Camera~3 
(WFC3/UVIS and WFC3/IR). Source lists from other instruments will
be added in the future, including the ACS High Resolution Camera
(ACS/HRC) and the Near Infrared Camera and Multi-Object Spectrometer
(NICMOS).  Data from other cameras (e.g., the imaging modes of the 
Space Telescope Imaging Spectrograph; STIS) may also be added at a later date.

HSC Version~1 was constructed using HLA Data Release~8 (DR8) images
and source lists. These include \textit{HST} data that was public as of 2014
June~1 for WFC3, 2011 February~16 for ACS, and 2009 May~11 for WFPC2.
Future releases of the HSC will include more recent data (except for WFPC2, which
was removed from \textit{HST} during Servicing Mission~4).

\begin{figure}[t]
\centering
\includegraphics[type=\plotext,ext=.\plotext,read=.\plotext,width=\linewidth]{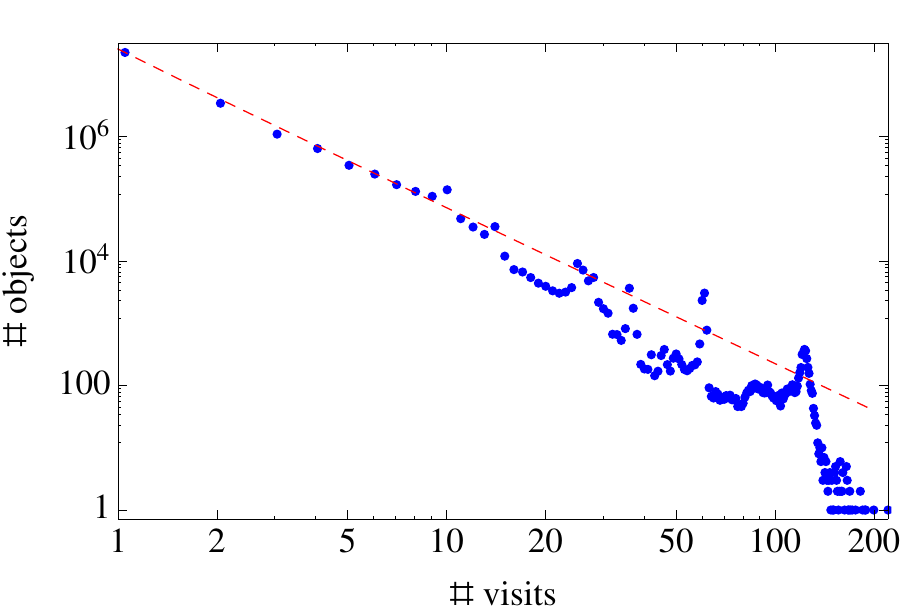}
\caption{The number of objects (matches) in the HSC catalog as a function of the number 
of visits in each match. The dashed line is a power law fit that falls off  as the number of visits 
to the power $-$2.5. Peaks in the distribution are partially due to repeat observations of: the 
Galactic bulge ($\sim$25 visits), M31 Halo ($\sim$60 visits), and M4 ($\sim$120 visits). 
\label{fig-visits}}
\end{figure}

Figure~\ref{fig-coverage} shows the patchwork nature of the \textit{Hubble}
observations, with only a small fraction (0.1\%) of the full sky being covered. 
This is a primary difference between the HSC and most other surveys and catalogs.  
The other major difference is the vast diversity in filter and exposure times  at 
different locations in the sky. Figure~\ref{fig-visits} is a log-log plot of the number 
of HSC catalog objects as a function of the number of independent visits to the 
object by \textit{HST}. The broad distribution is well fit by a power law fit that falls 
off as the number of visits to the power $-$2.5. 

\begin{deluxetable*}{lcccccc}
\tablecolumns{7}
\tablewidth{0pt}
\tablecaption{Basic HSC Statistics}
\tablehead{
\colhead{Instrument} & \colhead{\# Filters} &	\colhead{\# Images\tablenotemark{a}}	& \colhead{\# Detections\tablenotemark{b}} & \colhead{Area\tablenotemark{c}} & \colhead{Pixel Size} & \colhead{Aperture sizes\tablenotemark{d}}\\
\colhead{} & \colhead{} &\colhead{} & \colhead{} & \colhead{(sq deg)} & \colhead{(arcsec)}  & \colhead{(arcsec)} \\
\colhead{(1)} & \colhead{(2)} & \colhead{(3)} & \colhead{(4)} & \colhead{(5)} & \colhead{(6)} & \colhead{(7)}
}
\startdata
WFPC2     &  38 & 29,146    & $13\times10^6$ & 30    & 0.10 & 0.10, 0.30 \\
ACS/WFC   &  12 & \phn9,021 & $21\times10^6$ & 21    & 0.05 & 0.05, 0.15 \\
WFC3/UVIS &  47 & \phn4,772 & $31\times10^6$ & \phn5 & 0.04 & 0.05, 0.15 \\
WFC3/IR   &  15 & \phn6,763 & $14\times10^6$ & \phn5 & 0.09 & 0.15, 0.45 \\[3pt]
Total     & 112\phn & 49,702 & $79\times10^6$ &  61  & \nodata & \nodata
\enddata
\tablenotetext{a}{Number of images contributing source measurements.}
\tablenotetext{b}{Total number of source detections from all filters.}
\tablenotetext{c}{Sum of the areas of visit-level combined images.
This estimate ignores overlaps of these images.}
\tablenotetext{d}{Radii of the two circular apertures used for aperture photometry.}
\end{deluxetable*}

Table 1 provides some basic parameters and statistics for the
different instruments used in the HSC.  

Not all images within the Hubble Legacy archive are used by the HSC. About 35\% of the 
combined HLA images (filter-based combined images within a visit) were not included in 
the HSC for a variety of reasons. For example, the HSC does not include moving target 
images, images or source lists of low quality, images whose source lists contain less than 
10~sources, or images in visits that are likely affected by large numbers of cosmic rays
(i.e.,  exposures of more than 500 seconds with only one exposure taken for the visit).

The majority of the images are from the WFPC2, due to its longevity on \textit{HST} 
(16~years). An important difference between the WFPC2 and the later generation
ACS and WFC3 cameras is the larger WFPC2 pixel size (0.10~arcsec), again
highlighting the diversity of  \textit{HST} data.

Magnitudes based on observations from different instruments are reported in separate
columns of the HSC. For example, all three cameras have an F814W filter, with 
measurements appearing as the separate columns A\_F814W, W2\_F814W, and W3\_F814W.
In most cases users will analyze the data for the different instruments separately, but it is also 
possible to combine data together. In general this will require the use of photometric 
transformation equations and aperture corrections. This is discussed further in 
Section~\ref{subsec:ptsource_cross}.

\newpage\section{The Catalog}\label{sec:catalog}

The SExtractor software (Bertin \& Arnouts 1996) is used to produce the HLA source 
lists used in the HSC. Both aperture magnitudes (\texttt{MagAper1} and \texttt{MagAper2}: 
see Table~1; note that these are not total magnitudes since no aperture corrections have been 
made to these values), and total magnitudes (using the \texttt{MagAuto} algorithm in SExtractor 
to attempt to include all the light in the source) are provided in the HSC. The ABMAG system 
is used for the HSC; transformations are required to convert to other systems such as VEGAMAG 
or STMAG. DAOPHOT (Stetson 1987) source lists are also produced in the HLA, primarily for 
point sources. These are not used in the HSC however.

The radius used for the small and large aperture measurements (\texttt{MagAper1} and 
\texttt{MagAper2}) are 1~and 3~pixels for WFPC2 and ACS; 1.25 and 3.75 pixels for 
WFC3/UVIS; and 1.67 and 5~pixels for WFC3/IR. See Table~1 for the corresponding 
sizes in arcsec.  The sky background is defined as the median in an annulus from 5~to 
10~pixels.  In most cases, the detection threshold is set to three times the background 
noise, although it is adjusted in some regions in accordance with the source flagging 
(e.g., around very bright stars).

Unlike ACS and WFC3, WFPC2 source lists explicitly include a
correction for Charge Transfer Efficiency (CTE) loss, based on the
formulae from Dolphin (2009). Images with pixel-to-pixel corrections
using the algorithm developed by Anderson \& Bedin (2010) will be
used to construct ACS and WFC3 source lists in the future.

\subsection{How the Catalog is Constructed\label{sec-construction}}

The HLA source lists provide the starting point for the HSC
construction. These lists provide the characteristics of sources
that are contained in visit-based \textit{HST} images. The HSC construction
involves cross-matching the sources in these source lists. In broad
outline, the relative astrometry of overlapping images is improved
from the currently available HLA astrometry. Next, the sources that
are in the overlapping images are cross identified on the basis of
source position.  Aspects of the reduction pipeline, the astrometric
and cross-matching algorithms, and the properties of the Beta version
catalog, are described in Budav{\'a}ri \& Lubow (2012). See also
Budav{\'a}ri \& Szalay (2008) for a discussion of the Bayesian approach
at the heart of the cross-matching step, Lubow \& Budav{\'a}ri (2013)
for more details about the cross-matching algorithm, and Whitmore
et~al.\ (2008) for details about the early source list generation.

The basic steps involved in the construction of the HLA source
lists, and the subsequent construction of Version~1 of the HSC, are
briefly described below. More detailed descriptions of various
aspects of the process are available in the references provided
above, or in the HLA and HSC Frequently Asked Questions 
(FAQs; see Appendix~\ref{sec:appendix_D}).

\begin{enumerate}
\item Within the HLA, combine exposures for each filter within a visit using
\texttt{multidrizzle} (Fruchter 2009) for WFPC2 and ACS, and using 
\texttt{astrodrizzle} (\url{http://drizzlepac.stsci.edu/}) for WFC3. The 
background is subtracted for the WFPC2 and ACS, but not for the WFC3 
images. Inclusion of the background allows the source-finding parameters to 
be set to find fainter sources for the WFC3. In addition, a different spatial filter 
technique is used for sparse (Gaussian) and crowded (Mexican hat) fields in 
the WFC3 images (see \url{http://hla.stsci.edu/hla_faq.html#Source6} 
for a more detailed discussion).  New WFPC2 and ACS images will be made 
for the HLA using the approach used for the WFC3 in the future. For the 
current WFPC2 and ACS images, but not for the current WFC3 images,
a comparison is made with the SDSS and GSC2 catalogs to improve the
absolute astrometric positions of the HLA source lists at this stage. 
Additional astrometric adjustments are made at various other steps, as 
described below.

\item Combine the filter-based images into a ``white-light'' image (i.e., 
combine data from different filters, but within the same visit, to 
provide a deeper image with a wider wavelength range). A
white-light image serves as the detection image for the visit. No shifts are 
made to align WFPC2 and ACS exposures within a visit before combining
the data. For WFC3, an early version of the {\tt tweakreg} algorithm
within the \texttt{astrodrizzle} software package was used to align the
sub-images within a visit, prior to combining the images for the different filters.

\item Run SExtractor on the white-light (detection) images to obtain
white-light source lists. The filter-based source characteristics are then 
determined at each of the detection-based source positions. If no source 
is detected for a particular filter at the source position, then that 
information is maintained as a filter-based nondetection. Nondetections 
are  listed in the HSC Detailed Catalog (i.e., as ``N'' in the Det column), 
which is discussed in point~8 below. While both SExtractor and DAOPHOT 
source lists are available in the HLA, only the SExtractor source lists are used 
in the HSC. However, a comparison is made between various aspects of the 
DAOPHOT and SExtractor source lists at this stage to help weed out bad 
images and bad source lists before including them in the HSC database.
Appendix~\ref{sec:appendix_C} shows examples of the parameter files used to
produce the SExtractor and DAOPHOT source lists, in this case for WFPC2. 

Two different magnitudes are included in the HSC: \texttt{MagAper2} (aperture
magnitudes---see Table~1 for sizes on the sky) and \texttt{MagAuto} (SExtractor
estimates of the total magnitude primarily designed for extended sources). Smaller 
aperture measurements (\texttt{MagAper1}) can be recovered via the Concentration 
Index (CI), which is the difference between \texttt{MagAper1} and 
\texttt{MagAper2}.  Source properties such as the CI and the distance to nearby 
bright stars are also used to determine whether a source is likely to be an artifact. 
Such false detections are not included in the HSC. The complete set of attributes 
measured by SExtractor is accessible through links from the HSC back to the
original HLA source lists.

\item Apply astrometric ``pre-offsets'' based on cross matching with three reference 
catalogs: PanSTARRS, SDSS, and 2MASS. This is needed, for example, to reduce 
the typical $\sim$1--2 arcsec absolute astrometric errors for \textit{HST} images 
before Guide Star Catalog~2 became available in late 2005, to less than 0.3~arcsec. 
This step determines the statistical mode of the binned astrometric offsets between 
the \textit{HST} sources and those in a reference catalog. The method is very robust 
to large offsets and has a precision of a few tenths of an arcsec. Although this 
astrometric accuracy is not adeqaute for cross matching sources in different 
\textit{HST} images, it is sufficient for permitting convergence of the high accuracy 
Budav{\'a}ri and Lubow (2012) relative astrometry determination. Without the 
pre-offsets, the number of false matches across \textit{HST} source lists in very 
crowded fields (e.g., globular clusters) prevents the Budav{\'a}ri and Lubow (2012) 
relative astrometry correction algorithm from converging in many cases.

\item Separate the white-light images into groups of overlapping images.  Within 
each image group, determine the relative shifts (and rotations) needed to align 
the various images (see the Budav{\'a}ri \& Lubow 2012 paper for details on these 
two operations).  This reduces the relative astrometric accuracy from a few tenths of 
an arcsec to less than 10~mas in most cases (see Figure~\ref{fig-astrometry-offsets}
and the discussion in Section~4.2). Apply these image shifts to the sources in each 
white-light source list.

\item Cross-match the white-light sources by position. This is initially done using a 
friends-of-friends (FoF) algorithm with a specified search radius. FoF cross-matching 
can result in long chains of loosely connected sources that should be further split. 
Various ways of splitting each FoF match are considered by the software to determine 
the best partitioning.  Estimates of the astrometric uncertainties and the quality of a 
particular partioning for a match are determined by computing a Bayes factor using 
the formalism described in Budav{\'a}ri \& Szalay (2008).  We apply a greedy algorithm 
that reduces the number of partitions examined, as described in Budav{\'a}ri \& Lubow 
(2012). In most matches, the initial FoF match is found to have the best Bayes factor, 
and so no splitting is done.

\begin{figure}[t]
\centering
\includegraphics[type=\plotext,ext=.\plotext,read=.\plotext,width=\linewidth]{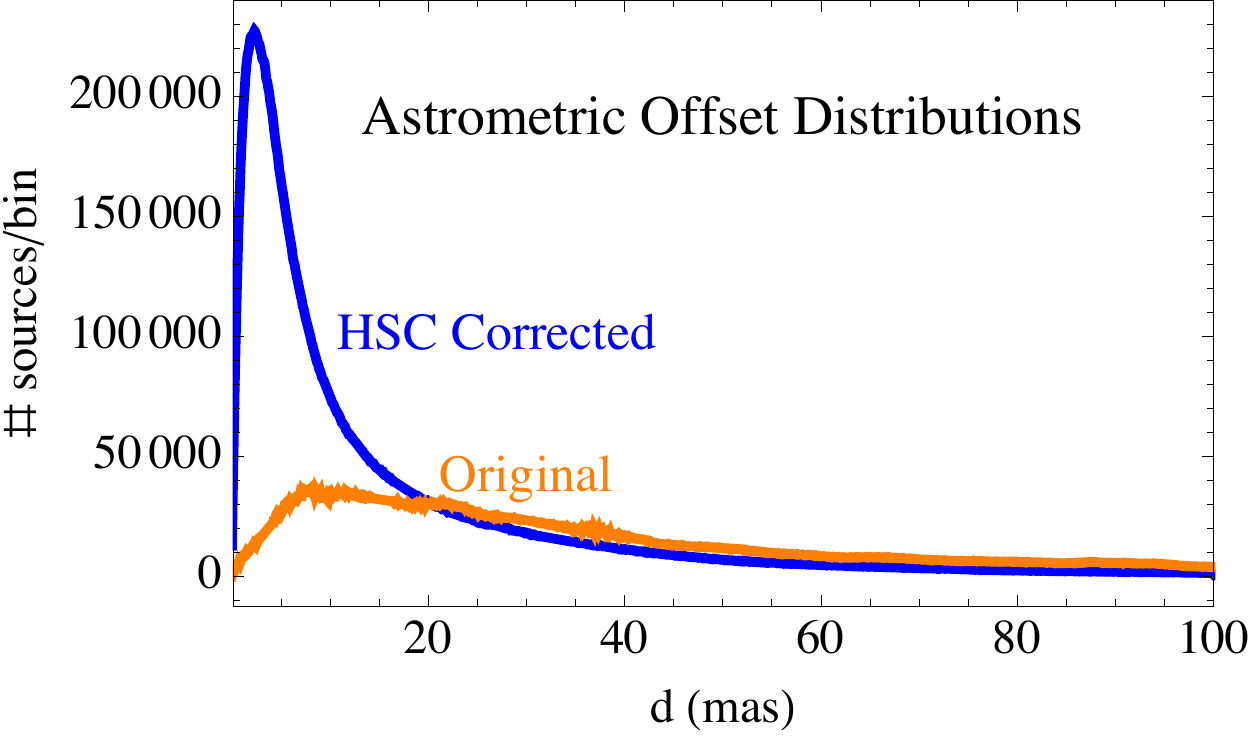}
\caption{Distribution of astrometric residuals before and after the Budav{\'a}ri \& Lubow 
(2012) algorithms are employed.  The areas under the two curves are the same, but the 
residual distribution before corrections has a very long tail that extends to much larger 
values.  The mode (peak) for the corrected curve is 2.3~mas.
\label{fig-astrometry-offsets}}
\end{figure}

\item Readjust the absolute astrometry for each group using PanSTARRS, SDSS, or 
2MASS as the reference. The absolute astrometric accuracy for the three reference 
catalogs is approximately 0.1~arcsec (e.g., see Pier et~al.\ 2003 for SDSS and Skrutskie 
et~al.\ 2006 for 2MASS; PanSTARRS uses SDSS as its astrometric backbone so should 
have similar accuracy). Hence the typical absolute astrometric accuracy for the HSC 
should be about 0.1~arcsec (but see Section~4.2.3 for results that show it may be 
somewhat better in some cases).

Figure~\ref{fig-coverage} shows that in 14\% of the images,  involving 32\% of the visits, 
there are insufficient matches with PanSTARRS, SDSS, or 2MASS to allow a correction. 
Most commonly this happens because both SDSS and PanSTARRS have limited sky 
coverage and 2MASS sources are sparse in the extragalactic sky. There are other reasons 
why absolute astrometric corrections cannot be made for parts of the HSC, for example 
for many far UV images when there are no objects in common with the three reference 
catalogs.  Another reason that there may be no absolute correction is that singleton 
images (images that do not overlap with other images---about 20\%) did not get 
corrected by the post-processing step in version~1. That will be corrected in future 
versions of the HSC. In the end, 80\% of the HSC matches have the Absolute 
Correction (\texttt{AbsCorr}) flag set to yes (Y).

\item Using the white-light source lists with corrected astrometry, the cross-matched 
white-light sources, and the filter-based source lists, build the HSC Detailed Catalog, 
which includes visit- and filter-based information about each source.  Also, build the 
HSC Summary Catalog. This contains the aggregate properties for each set of cross-matched 
sources, such as the mean position, the mean magnitudes and their standard deviations. It 
also includes some useful ancillary data such as the galactic extinction (i.e., $E$($B$--$V$) 
values from Schlegel 1998) at that position on the sky. 
\end{enumerate}

Steps 1 through 3 above use the HLA image and source list processing, while steps~4 
through~8 are carried out in a Microsoft SQL Server database.  The catalogs are made 
available though a set of database tables that can be accessed by various user interfaces, as 
will be discussed in \S5. In addition, there are database stored procedures and functions 
that use special indexes and algorithms to provide rapid access to HSC information. This 
makes it possible for users to make complex requests through the user interfaces, such as 
cross matching HSC against a user-supplied set of positions, with fast response times.
Currently, the Discovery Portal, the Hubble Legacy Archive image display, and the 
MAST (Mikulski Archive for Space Telescopes) search forms access the HSC database 
using these stored procedures and functions. In addition, the CasJobs interface allows 
users to directly access the functions and run more general SQL queries. See Appendix~B 
for more information about these four interfaces.

\section{Quality Assessment}\label{sec:quality}

A three-pronged approach is used to characterize the quality of the HSC. We first examine 
a few specific datasets, comparing magnitudes and positions directly for repeat measurements. 
The comparisons are first made using the same instrument and filter, and then made using
different instruments and filters.

The second approach is to compare repeat measurements for the full database. While this 
provides a better representation of the entire dataset, it can also be misleading since the tails 
of the distributions are generally caused by a small number of bad images, bad source lists, 
and other artifacts.

The third approach is to produce a few well-known astronomical figures (e.g., color-magnitude 
diagram for the outer disk of M31from Brown et~al. 2009) based on HSC data, and compare 
them with the original study.  Three examples of this third approach are provided in Appendix~A.

This three-pronged approach is hierarchal in nature: 1)~a spot check on the consistency and 
quality of the source lists for a few specific data sets, 2)~a check that the entire dataset is relatively 
homogenous and of high quality, and 3)~a check that we are consistent with completely
independent datasets or independent analysis techniques.

As stressed in other parts of this paper (e.g., Section 5, ``Caveats and Warnings''), it is important 
to keep in mind that parts of the HSC can be very non-uniform. Hence, researchers cannot 
assume that the results reported in this section represent the entire database. If uniformity is 
important for a specific science project, a careful examination of the data is required, including 
viewing the images themselves. In many cases it is possible to filter the HSC data and improve 
the uniformity of the data. This topic will be discussed in Section~4.1.4.

\subsection{Photometric Spot Checks}
\subsubsection{Point Source Photometry---Single Instrument/Filter Checks}\label{subsec:ptsource}

Since we are primarily interested in stellar photometry in this section, aperture magnitudes (i.e., 
\texttt{MagAper2}) are used throughout.

Our first photometry check examines the Brown et~al.\ (2009) deep ACS/WFC 
observations of the outer disk of M31 using objects within 2.5~arcmin of the J2000 search 
position $00^{\rm h}49^{\rm m}08.09^{\rm s}$ ${+}42\arcdeg44^\prime55.0\arcsec$.
The observing plan for this proposal (ID~= 10265) resulted in approximately 60~separate 
one-orbit visits (not typical of most \textit{HST} observations), hence providing an excellent 
opportunity for determining the true uncertainties by examining repeat measurements.

\begin{figure}[t]
\centering
\includegraphics[type=\plotext,ext=.\plotext,read=.\plotext,width=\linewidth]{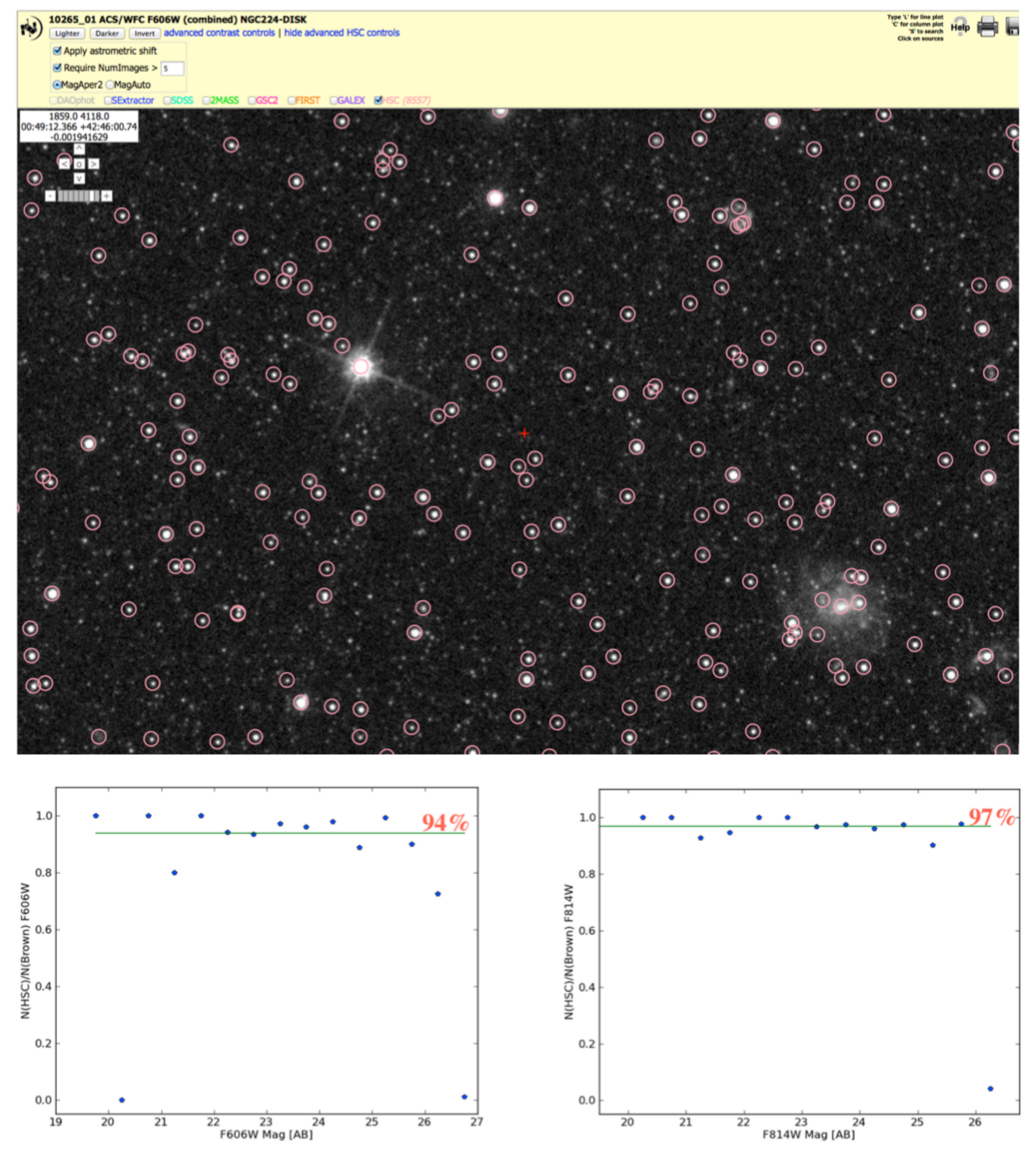}
\caption{Example of the quality of the HSC in the outer disk of M31overlaid on an ACS image 
from proposal 10265. The bottom plots show the completeness levels for the F606W and F814W 
observations when compared to the Brown et~al.\ (2009) study.
\label{fig-M31-image}}
\end{figure}

Figure~\ref{fig-M31-image} shows a small part of the field with the HSC overlaid. Only 
sources detected on more than five images (\texttt{NumImages} $>5$) are included in order 
to filter out cosmic rays. Note that relatively faint stars are not included; the more recent
WFC3 HLA sources lists (and future ACS and WFPC2 source lists) are more aggressive in 
this regard. The completeness limits for the HSC when compared to the Brown et~al.\ 
(2009) catalog are roughly 95\% in the F606W and F814W filters out to about 26th magnitude.

\begin{figure}[brown_2plot]
\centering
\includegraphics[type=\plotext,ext=.\plotext,read=.\plotext,width=0.8\linewidth]{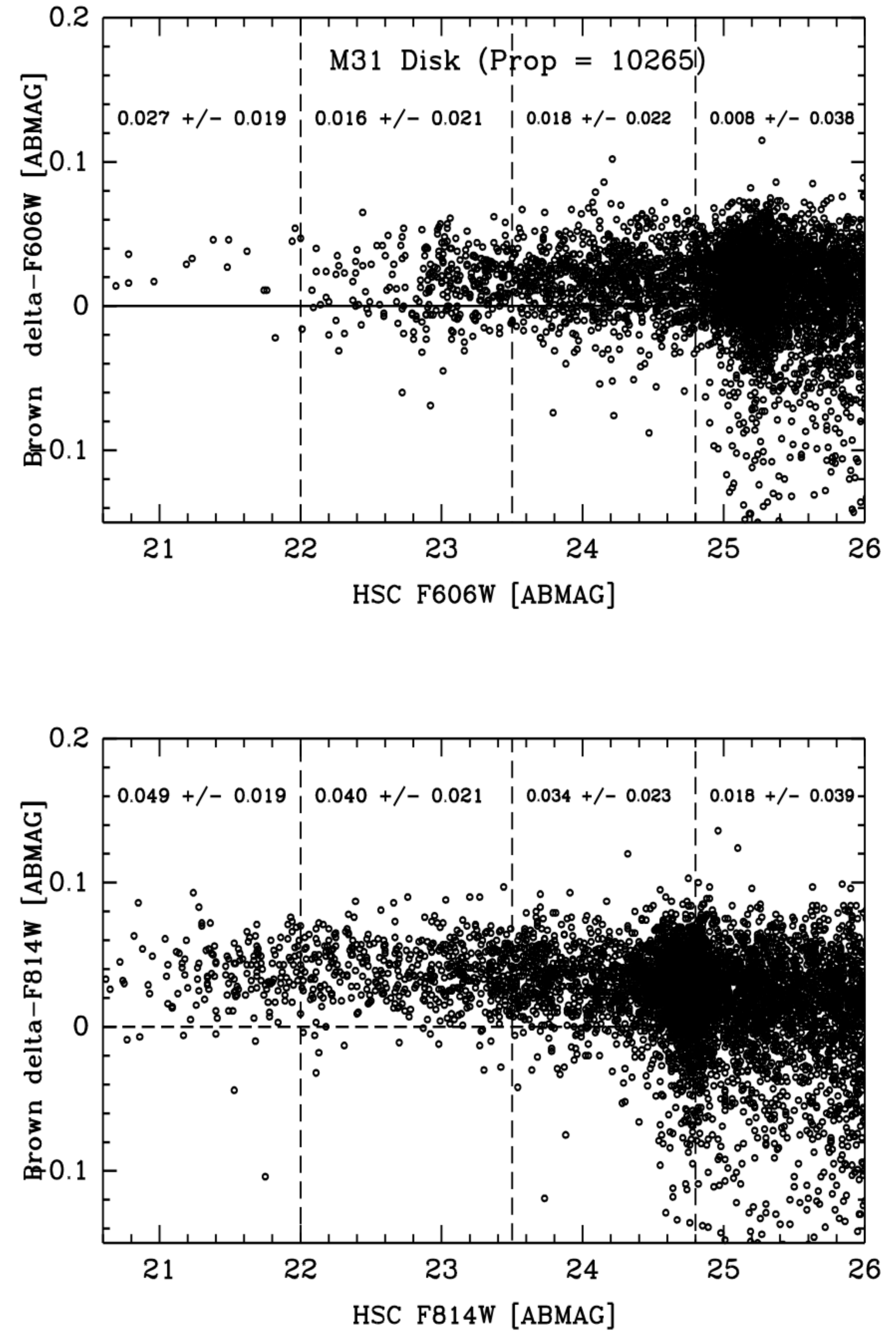}
\caption{Comparison of HSC and Brown et~al.\ (2009) photometry, with means 
and RMS scatter listed  for four magnitude ranges.  The top (bottom) panel shows 
the F606W (F814W) magnitude difference; objects are systematically slightly 
fainter in the HSC.
\label{fig-M31-photometry2}}
\end{figure}

Figure~\ref{fig-M31-photometry2} shows that the photometric agreement
between the HSC and Brown et~al.\ catalog is quite good, with zeropoint 
differences of only a few hundredths of a magnitude after corrections from 
ABMAG to STMAG and from aperture to total magnitudes are made. The 
photometric scatter is about 0.04~mag in general, and  better than 0.02~mag
for the brighter stars. The zero point offsets between the HSC and the Brown 
et~al.\ catalog are likely to be due mainly to the inclusion of a CTE (Charge 
Transfer Efficiency) correction by Brown et~al., but not by the HSC in 
Version~1. The sense of the difference is in the right direction, with the HSC 
magnitudes slightly fainter, and the magnitude of the offset is also reasonable, 
since Brown et~al.\ (2006) comment that the expected CTE corrections are a 
``few hundredths of a magnitude'' in fields like these.  More details are
available in HSC Use Case \#1 (see Appendix~A.1).

\clearpage
The use of short, one-orbit visits in proposal 10265 leads to one of the common limitations 
of the HSC, namely brighter completeness limit for HLA source lists than are possible by 
combining all the data. For example, the deep, co-added, 30 orbit for each filter image used 
by Brown et~al.\ goes roughly four magnitudes deeper than the HSC, as will be shown in \S4.4.1.

A more representative comparison would be with catalogs produced using one-orbit visits.
These cover a  wide range of  quality and completeness limits depending on the specific science 
requirements. Typical studies reach roughly the same level as the WFC3 source lists in the HSC
while state-of-art studies like the Panchromatic Hubble Andromeda Treasury program (PHAT; e.g., 
see Williams et~al. 2014) go roughly two magnitudes deeper in the confusion-limited IR images. A 
detailed comparison between the HSC and PHAT will be provides as a use case for the Version~2 
HSC release, currently planned for spring of 2016.  

\subsubsection{Point Source Photometry---Error Estimates}

\begin{figure}[t]
\centering
\includegraphics[type=\plotext,ext=.\plotext,read=.\plotext,width=0.8\linewidth]{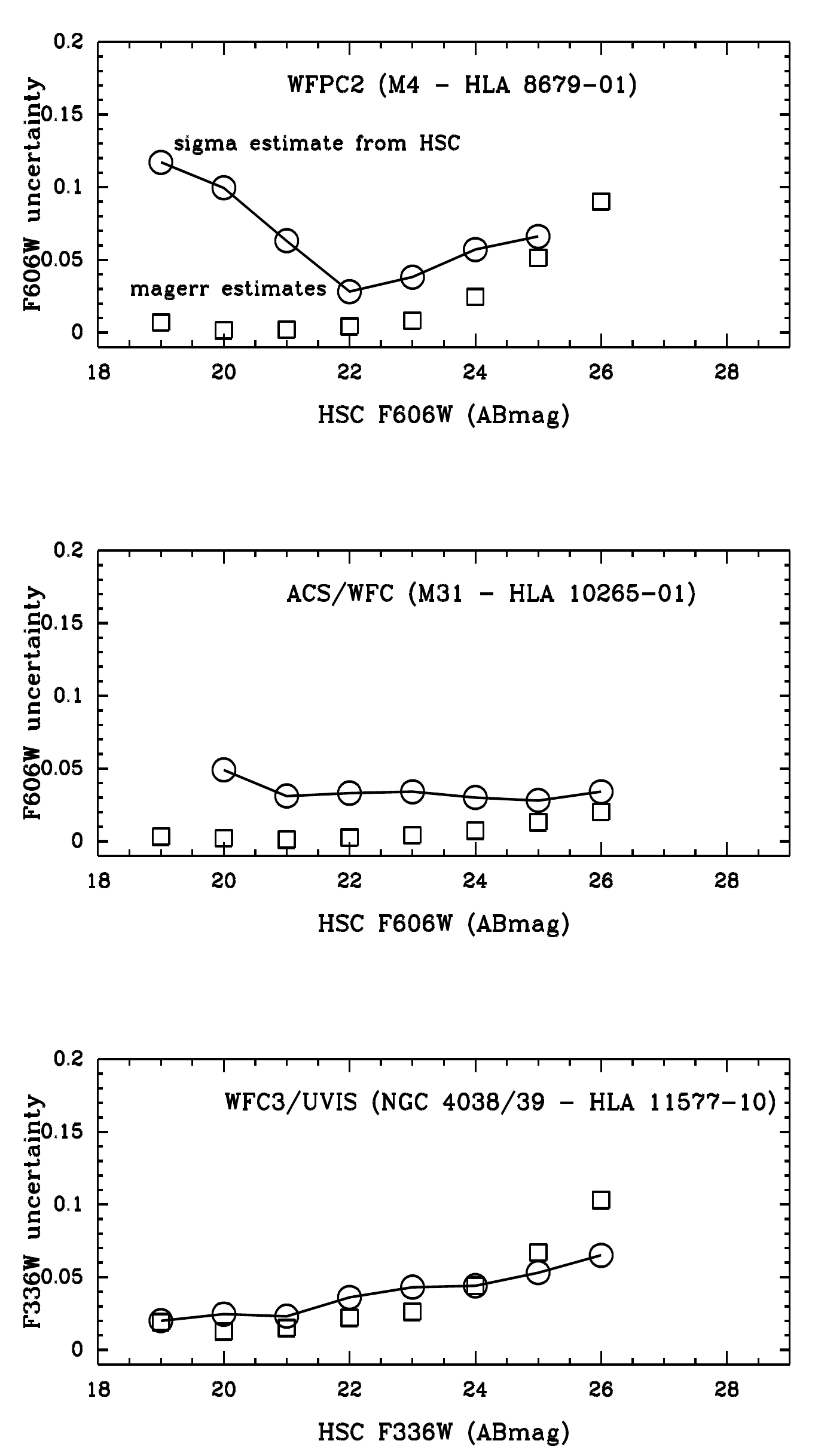}
\caption{Comparison of HSC sigma values (i.e., RMS scatter based on repeat measurements [circles]
with magerr estimates based on SExtractor squares). The upturn in the sigma estimate for the WFPC2 
at bright magnitudes is due to inadequate filtering of a few saturated stars, as discussed in the text.
\label{fig-photometry-sigma}}
\end{figure}

Figure~\ref{fig-photometry-sigma} shows a comparison between estimated photometric errors 
based on SExtractor measurements (i.e., magerr), and the true scatter based on repeat measurements 
(i.e., values of sigma reported in the HSC summary catalog). We find that the quoted values of 
magerr are roughly a factor of three too low for WFPC2 and ACS observations, but are in relatively 
good agreement for the newer WFC3 source lists.

We also note that the sigma estimates increase rather dramatically at bright magnitudes for 
WFPC2, and to a lesser degree for ACS as well. This is due to the inclusion of a few saturated 
stars that have made it through the filtering designed to flag and remove them (see the discussion 
in \S4.1.4). These problems will be rectified in the near future when the pipeline developed for the 
newer WFC3 source lists is used to produce the next generation of WFPC2 and ACS HLA source lists.

\begin{figure}[t]
\centering
\includegraphics[type=\plotext,ext=.\plotext,read=.\plotext,width=1.0\linewidth]{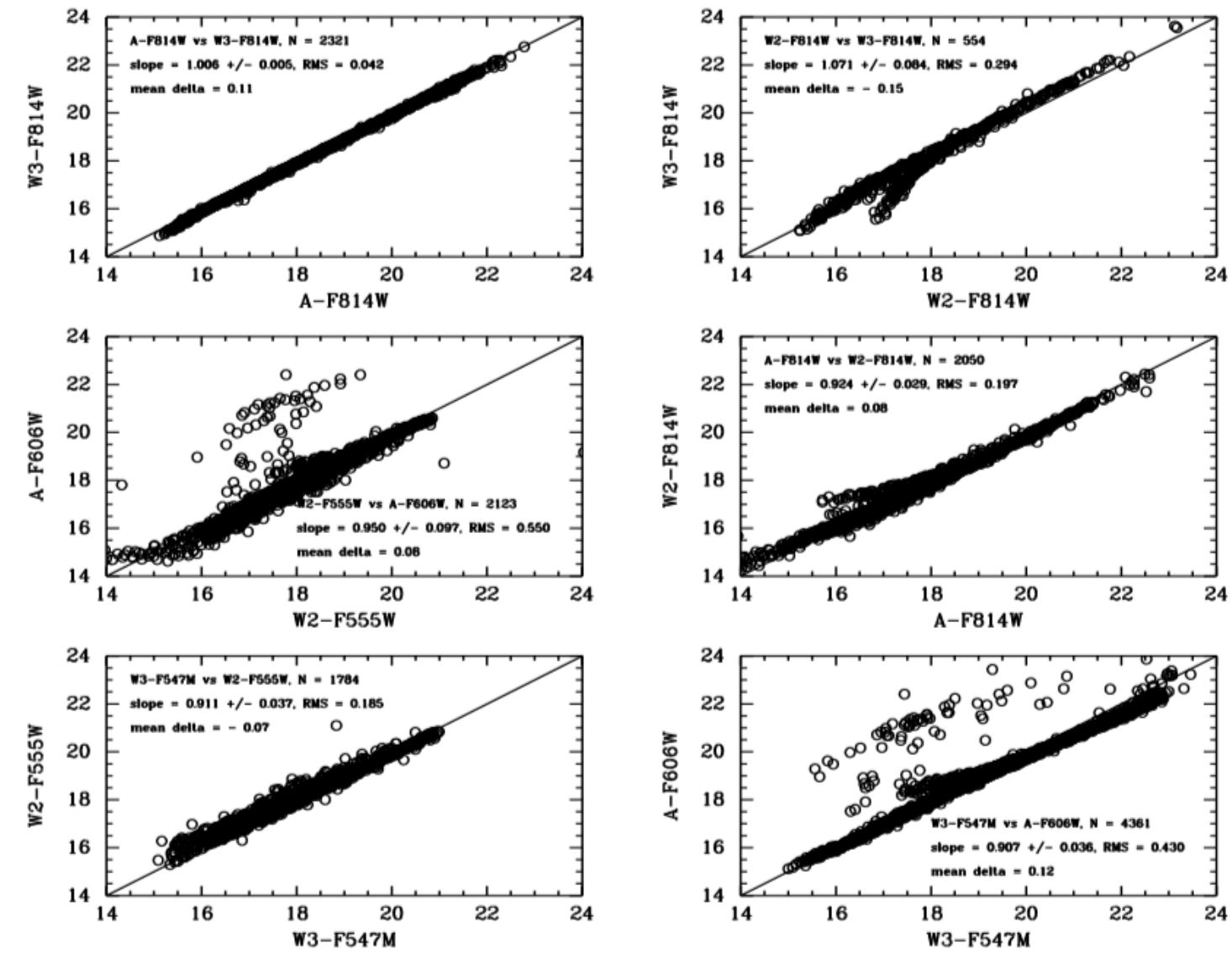}
\caption{Comparisons of repeat measurements for similar filters in the globular cluster M4. Note that 
photometric transformations between the instrument/filter combinations would be required before the 
different observations could be combined, if desired.
\label{fig-M4-repeats}}
\end{figure}

\begin{figure}[t]
\centering
\includegraphics[type=\plotext,ext=.\plotext,read=.\plotext,width=1.0\linewidth]{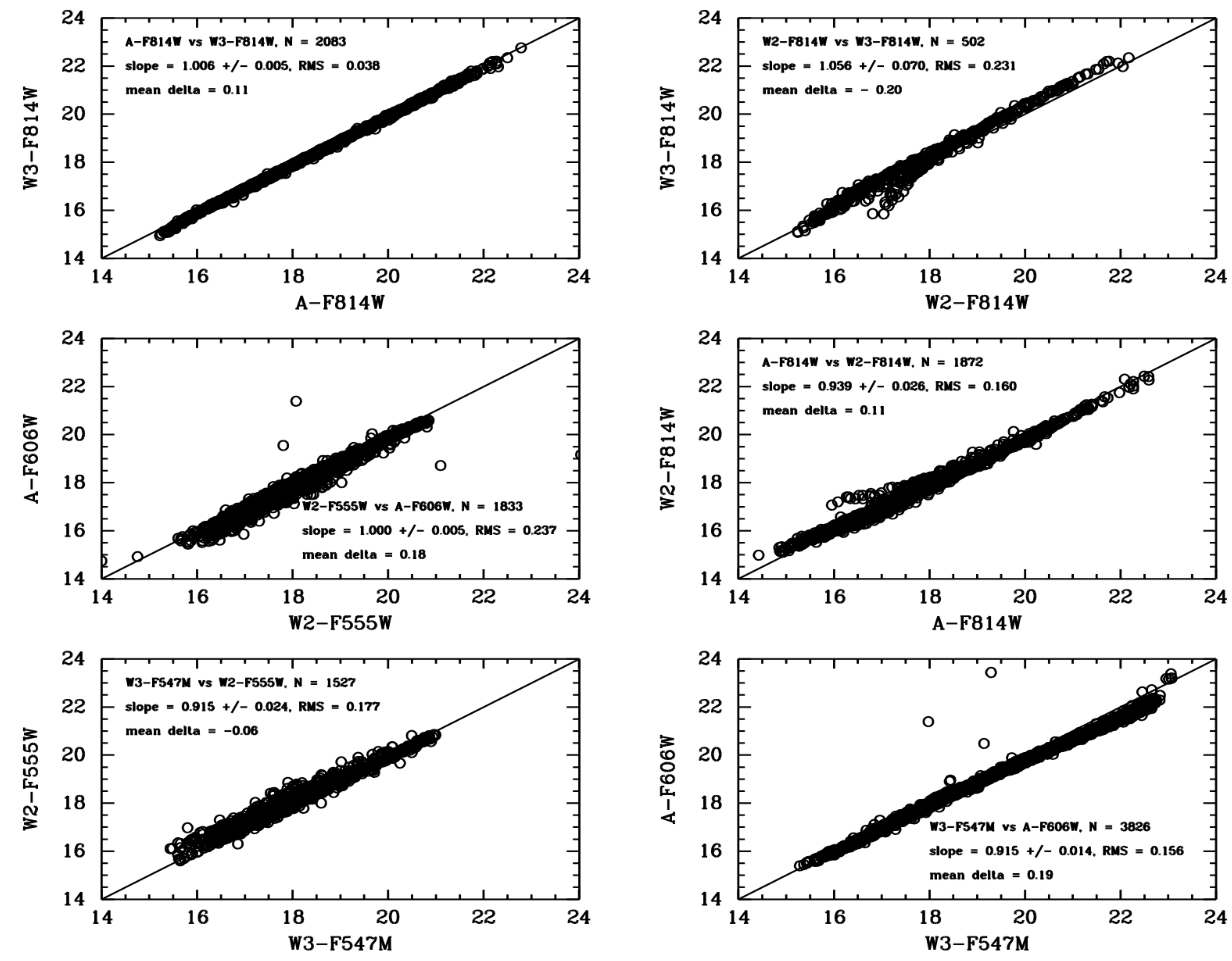}
\caption{Same as Figure~\ref{fig-M4-repeats}, but with the four constraints discussed in 
Section~4.1.4 imposed.
\label{fig-M4-repeats-clean}}
\end{figure}

\begin{figure}[t]
\centering
\includegraphics[type=\plotext,ext=.\plotext,read=.\plotext,width=1.0\linewidth]{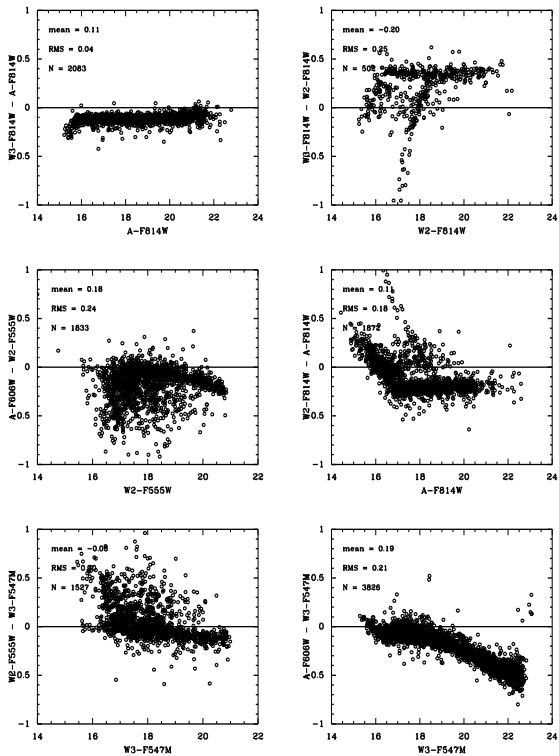}
\caption{Same as Figure~\ref{fig-M4-repeats-clean}, but with residuals rather than a one-to-one 
plot to show the small-scale features.
\label{fig-M4-repeats-rms}}
\end{figure}

\subsubsection{Point Source Photometry---Cross-Instrument/Filter Checks}\label{subsec:ptsource_cross}

The globular cluster M4 (using a search within $200\arcsec$ of 
$16^{\rm h}23^{\rm m}38.66^{\rm s}$ ${-}26\arcdeg32^\prime10.9\arcsec$) provides a good 
opportunity to compare the HSC photometric system for all three instruments. 
Figure~\ref{fig-M4-repeats} shows comparisons in the ``V'' filters (i.e., WFPC2-F555W, ACS-F606W, 
and WFC3-F547M) and ``I'' filters (i.e., WFPC2-F814W, ACS-F814W, and WFC3-F814W).

Starting with the best case, ACS-F814W vs.\ WFC3-F814W shows excellent results, with a slope 
near unity, values of RMS around 0.04 magnitudes, and essentially no outliers. The good 
agreement also suggests that ACS-F814W and WFC3-F814W measurements can be added 
together with little loss of photometric accuracy. This is not true, as we will see below, when the 
filter bandpasses are not as similar. In general, photometric transformations are necessary before 
combing observations using different instruments.

An examination of the WFPC2-F814W vs.\ WFC3-F814W and ACS-F814W vs.\ WFPC2-F814W 
comparisons show that there is an issue with the WFPC2 data.  The short curved lines deviating from 
the one-to-one relationship show evidence of the inclusion of a small number of slightly saturated
measurements for bright stars (i.e., roughly 5\% of the data), as already mentioned in the discussion of 
Figure~\ref{fig-photometry-sigma}.

 There is a similar but smaller issue with the  ACS data, as shown by 
the WFC3-F547M vs.\ ACS-F606W comparison. 

Much larger deviations are also seen in the two panels making use of ACS-F606W observations, where 
a cloud of outliers is found several magnitudes off the one-to-one line. These are caused by combining
data from short (20~sec) and long (1800~sec) sub exposures. These issues will be fixed in future versions 
of the HSC, but it is also relatively easy to filter them out, as discussed in \S4.1.4.

A careful look at Figure~\ref{fig-M4-repeats} also shows systematic deviations in the slope of the 
relationships, with deviations of a few tenths of a magnitude at the extremes (e.g., WFC3-F547M vs.\ 
ACS-F606W). Figure~ \ref{fig-M4-repeats-rms} in \S4.1.4 shows this more clearly.  These are 
examples where the filters are not well matched (e.g., the central wavelength and width are 591.8 
and 158.3~nm for the ACS-F606W filter but 544.7 and 65.0~nm for the WFC3-F547M filter). 
Hence sources with different colors (and hence different brightnesses since this is a globular cluster 
with a well-defined main sequence) deviate in the two filters. A photometric transformation would 
need to be made before photometry in these two filters could be combined. The comparison is made 
here in order to evaluate the RMS scatter, not to imply that the data from different instruments/filters 
can be added together without the loss of a few tenths of a magnitude in accuracy.

Other complications that can cause deviations are issues having to do with Charge Transfer Efficiency 
(CTE) loss (i.e., for WFPC2 a correction is made using the Dolphin 2009 formula, but no corrections
are made for ACS and WFC3 in Version~1), differences in aperture corrections (typical differences 
between the different instruments are about 0.1~mag for the ACS, WFC3/UVIS, and WFPC2), and 
differences in exposure times (e.g., resulting in different completeness limits and signal-to-noise 
ratios---see the transition at about the 19th magnitude in the WFPC2-F555W vs.\ ACS-F606W 
diagram with larger scatter at brighter rather than fainter magnitudes).

\subsubsection{Filtering out Artifacts}\label{sec:filtering}

As stressed throughout this paper, the diverse nature of the \textit{HST} archival database can result 
in a number of artifacts. However, we also note that the availability of multiple observations in many
cases provides the opportunity to identify artifacts and filter them out, a circumstance that is not always 
possible with more limited datasets where similar artifacts may still be present but go undetected.

Figure~\ref{fig-M4-repeats-clean} shows the same comparisons as Figure~\ref{fig-M4-repeats}, but 
with four constraints included. These are:
\begin{itemize}
\itemsep0em

\item \texttt{NumImages} $ > 2$ (to remove residual cosmic rays),

\item CI $<1.4$ (to remove extended sources and blends),

\item \texttt{CI\_Sigma} $<0.5$ (to remove partially saturated stars), and

\item \textit{filter}\texttt{\_Sigma} $ < 0.2$ (to remove low S/N data and saturated stars),
\end{itemize}
\noindent where CI is the Concentration Index (i.e., the difference between the small and large 
aperture magnitudes---see Table 1), and \texttt{CI\_Sigma} and \textit{filter}\texttt{\_Sigma} 
refer to the RMS scatter among repeat measurements of the CI value and the magnitude in a given
filter.

As shown in Figure~\ref{fig-M4-repeats-clean}, the number of artifacts and discrepant points is 
greatly reduced, with only 3/3826 (0.1\%) artifacts remaining in the WFC3-F547M vs.\ ACS-F606W 
comparison with residuals greater than 1~mag. The values of RMS scatter from the line are also reduced, 
in some cases by more than a factor of two. The values of ``true RMS'' shown in 
Figure~\ref{fig-M4-repeats-clean}, which are the values after the remaining outliers have been removed,
range from 0.04 to 0.17 magnitudes.

Figure~\ref{fig-M4-repeats-rms} shows a version of Figure~\ref{fig-M4-repeats-clean} with residuals 
rather than a one-to-one plot. This allows us to see the small-scale features more clearly, especially the 
slight curvatures due to the mismatch in filters (e.g., the bottom right panel) and the increase in scatter for 
shorter exposures (e.g., the two bottom left panels), as discussed in \S4.1.3 .  These figures also show the 
inherent danger of combining data from different instruments and filters. Although this might be 
appropriate in certain cases (e.g., the upper left panel), it should only be attempted with great care. 

While the specific criteria may change for different datasets and scientific purposes, some combination of the 
four artifact filters employed in this section can often be used to improve the photometry from the HSC.

\begin{figure}[t]
\includegraphics[type=\plotext,ext=.\plotext,read=.\plotext,width=\linewidth]{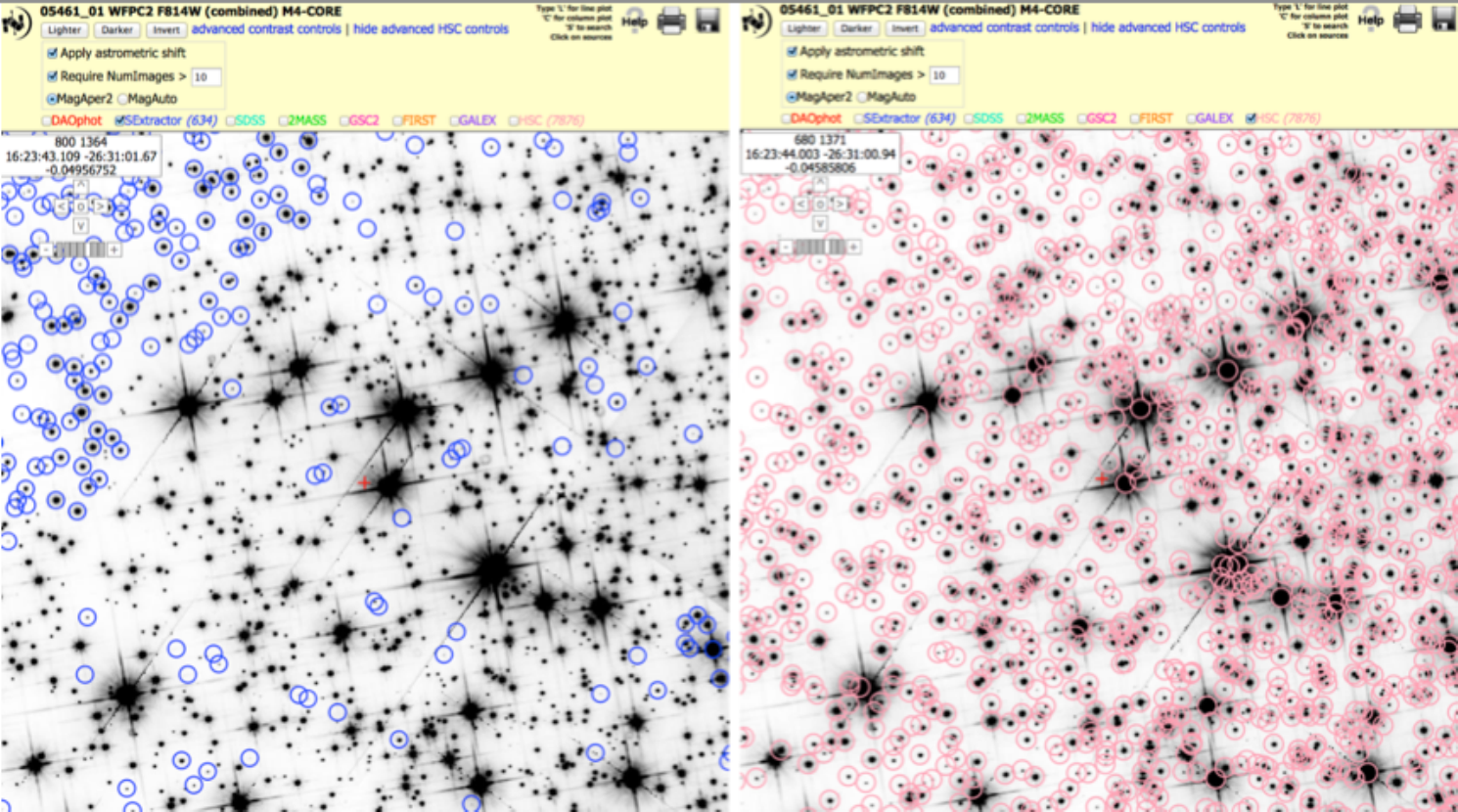}
\caption{The left image shows a WFPC2 SExtractor source list (blue) in the globular cluster M4. Note 
how nonuniform the coverage is with missing sources where the background is high. The right image
shows the more uniform HSC coverage (pink circles). It is more uniform due to the presence of WFC3 
source lists in this field. See discussion in \S4.1.3 for more details. 
\label{fig-M4-images}}
\end{figure}

Another form of ``artifact'' is the non-uniformity inherent in a dataset as diverse as the \textit{Hubble} 
archives.  This is accentuated by the current poorer quality of the WFPC2 and ACS HLA source lists
relative to the more recently generated WFC3 source lists.  For example, Figure~\ref{fig-M4-images} 
shows that many sources are missed in regions with high background in this WFPC2 image. SExtractor 
tends to combine high background regions into larger ``extended" objects, missing obvious stars in cases 
like Figure~\ref{fig-M4-images}. While SExtractor parameters can be tuned to largely alleviate this 
problem for a specific image, this is difficult for the HLA due to the diversity of the images. As discussed in 
the first bullet in \S3.1, a Mexican hat kernel is now used in the HLA for WFC3 images in crowded regions, 
largely eliminating this problem. The same algorithm will be used for WFPC2 and ACS in the future. 

While the overall coverage of the HSC is quite good (i.e., the pink circles in Figure~\ref{fig-M4-images}), 
thanks mainly to the WFC3 images in this region, users should keep in mind that just because a given 
observation is missing in the HSC does not mean that it has not been observed by \textit{Hubble}.

More details about the comparisons discussed above, as well as other examples relevant to photometric 
accuracy, can be found in HSC Use Case~\#1 (Stellar Photometry in M31), HSC Use Case~\#2 (Globular 
Clusters in M87 and a Color Magnitude Diagram for the LMC), and HSC Use Case~\#5 (White Dwarfs 
in the Globular Cluster M4).  See Appendix~\ref{sec:appendix_D} for URLs for these and other HSC
use cases.

\begin{figure}[t]
\centering
\includegraphics[type=\plotext,ext=.\plotext,read=.\plotext,width=\linewidth]{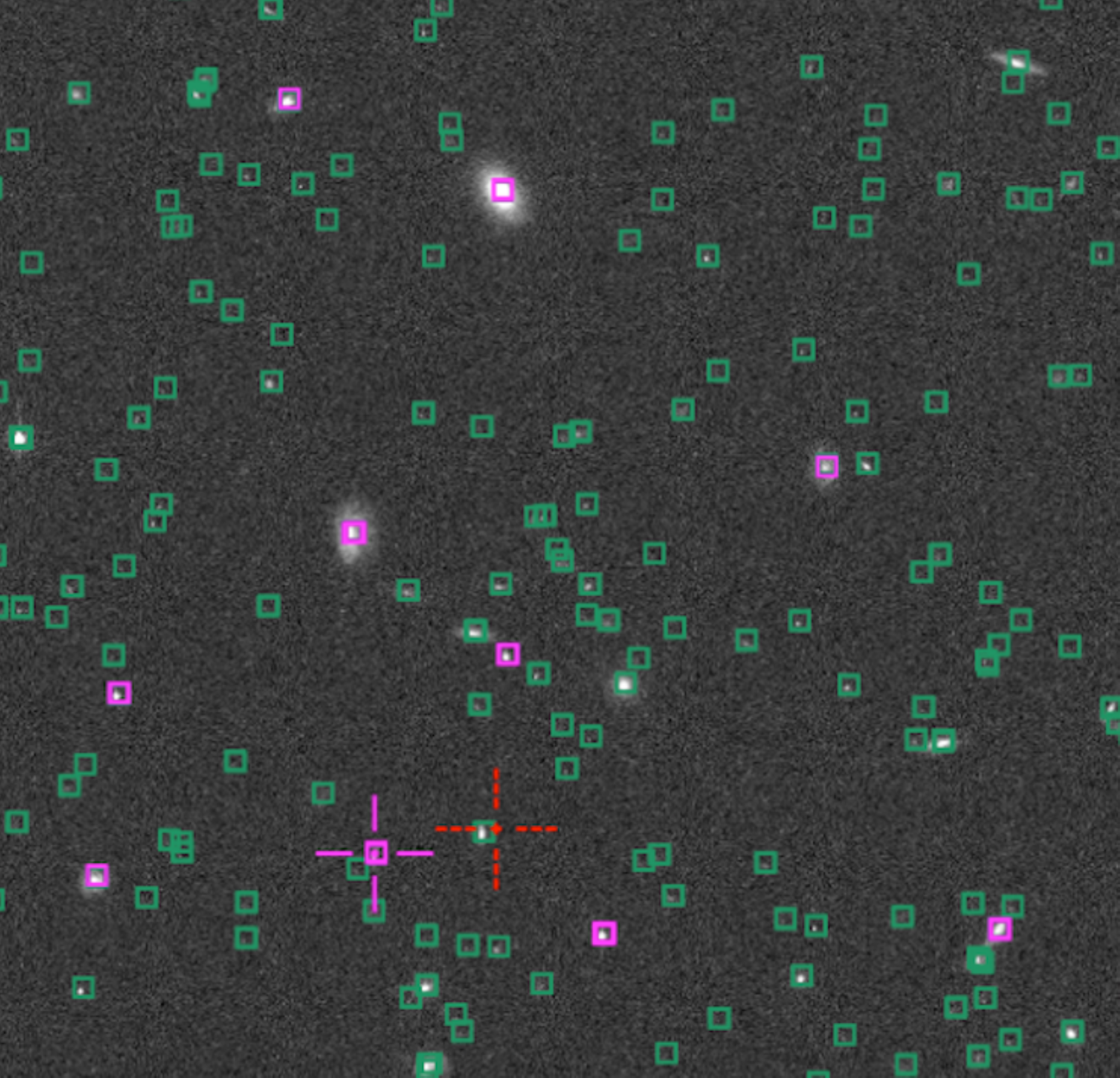}
\caption{Comparison of HSC (green) and SDSS (pink) in a small part of the Hubble Deep Field. This shows 
the increased depth using the HSC (approximately 150 objects) compared to the SDSS (9~objects). The image 
was made using the MAST Discovery Portal with an \textit{HST} image as the background.
\label{fig-HDF}}
\end{figure}

\begin{figure}[t]
\centering
\includegraphics[type=\plotext,ext=.\plotext,read=.\plotext,width=1.0\linewidth]{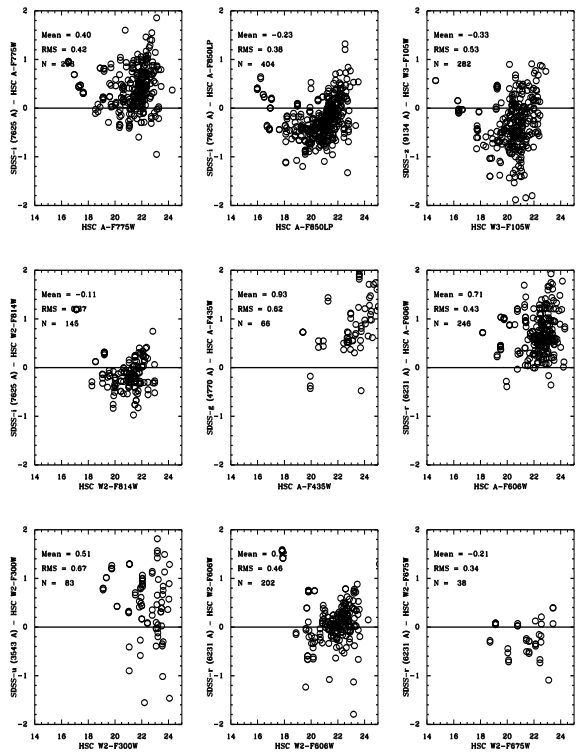}
\caption{Comparison between HSC photometry (\texttt{MagAuto}) and SDSS (DR12, Alam et~al.\ 2015; 
psfMag values) photometry for the Hubble Deep Field (i.e., RA~= 189.206, DEC~= 62.2161, $r=500$ arcsec). 
Note that photometric transformations between the instrument/filter combinations would be required before 
the different observations could be combined, if desired.
\label{fig-SDSS-photometry}}
\end{figure}

\subsubsection{Extended Object Photometry---SDSS Observations in the Hubble Deep Field}

In this section we make photometric comparisons with extended targets, such as distant galaxies. Hence, 
values obtained using the Source Extractor algorithm \texttt{MagAuto} are used to estimate the total
magnitudes rather than the aperture magnitudes (\texttt{MagAper2}) used in the previous sections.

The Sloan Digital Sky Survey (SDSS) has been tremendously successful, due to both the high quality, 
wide-field, uniform database, and to the extensive extraction and analysis tools it has made available
to researchers. It has taken the field of ``database astronomy'' to a new level, and in many ways is the 
inspiration for the HSC.

A comparison between the HSC and SDSS provides an opportunity for highlighting both the similarities 
(e.g., agreement between photometric results; availability of CasJobs) and differences (e.g., the HSC
goes deeper but with ``pencil beam'' coverage; the HSC can be very non-uniform in certain regions).

Figure~\ref{fig-HDF} shows the overlap between the HSC and SDSS coverage in a small part of the 
Hubble Deep Field (HDF).  There are 9~objects in common out of the roughly 150 HSC sources in this
field. The SDSS (using DR12; Alam et~al. 2015) has a completeness limit around $r=22.5$ mag while
the HSC goes to ACS-F775W~$= 26.0$ mag.

Figure~\ref{fig-SDSS-photometry} shows the photometric comparisons between the HSC and SDSS 
for a wide variety of filters.  We find reasonably good agreement. Both the scatter and the offsets are
typically a several tenths of a magnitude. The offsets primarily arise from differences in photometric 
filters and bandpass, since no transformations have been made for these comparisons.

The best agreement is between WFPC2-F814W and SDSS-i,  with a mean offset of $-$0.11 and a RMS
scatter of 0.37 mag. The mean photometric scatter for repeat HSC measurements in the ACS-F850LP
filter is about 0.10 mag while for the SDSS-i the scatter is about 0.15 mag. In quadrature these add to 
0.18 mag, explaining some but not all of the observed scatter in the comparison. Differences between
SExtractor parameters (e.g., thresholds, filtering algorithms, and deblending values) in the HSC and 
SDSS are responsible for most of the rest of the difference.

The relatively small scatter for this particular comparison reflects the fact that these two filters are very 
similar, hence the transformation is nearly one-to-one.  This is not true for many of the other comparisons
in Figure~\ref{fig-SDSS-photometry}.

Figures~\ref{fig-HDF} and~\ref{fig-SDSS-photometry} used \texttt{NumImages} $> 10$, and a value for 
the cross-match radius of 0.2~arcsec. Care must be taken to choose optimal values for these parameters to filter
out artifacts and mismatches between sources. Even so, some manual weeding is often necessary. In this 
particular case the objects are isolated enough to make this a minor issue, with only a few mismatches
present (i.e., the outliers in the SDSS-r vs.\ HSC ACS-F606W comparison).

\subsection{Astrometric Spot Checks}

The quality of point source astrometry for the HSC can vary for reasons similar to those relevant to photometric 
measurements. These include non-uniformities due to the wide range of instruments, different exposures times, 
and different observing strategies used by the observers. In this section we use the same approach as employed for 
photometry, starting with comparisons using a single instrument, then comparing different instruments, and latter 
making comparisons for the entire database.

\begin{figure}[t]
\centering
\includegraphics[type=\plotext,ext=.\plotext,read=.\plotext,width=\linewidth]{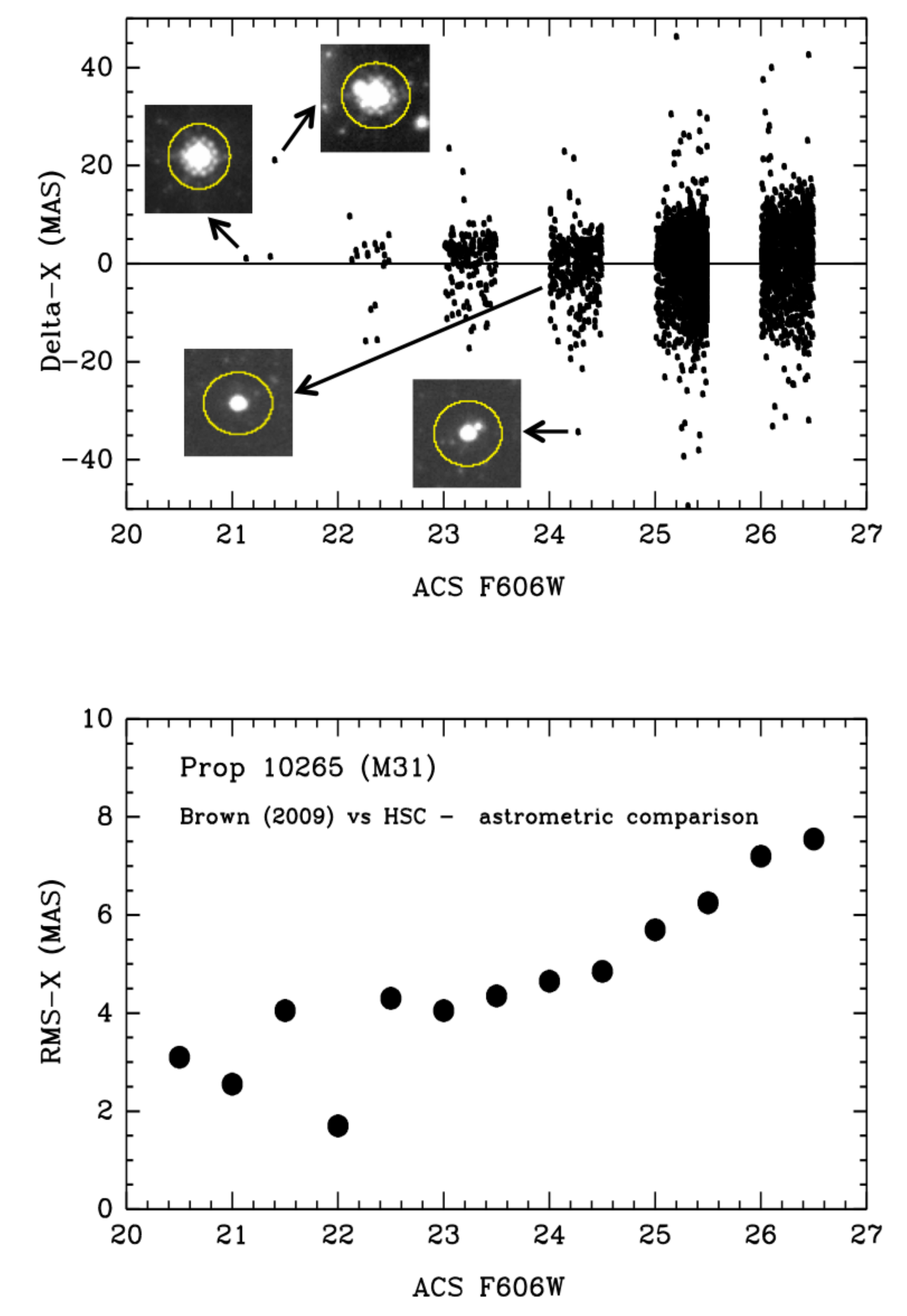}
\caption{Comparison of positions (in the X direction) between the HSC and the Brown et~al.\ (2009) HLSP 
catalog. The top panel shows comparisons for each object while the bottom panel shows the RMS scatter in repeat 
measurements as a function of magnitude. 
\label{fig-Brown-astrometry}}
\end{figure}

\subsubsection{Relative Point Source Astrometry---Single Instrument (ACS/WFC) }

We begin with the same dataset used in \S4.1.1, namely the Brown et~al.\ (2009) ACS/WFC observations in 
the outer disk of M31 (prop ID~= 10265; see HSC Use Case~\#1, and for more details, Archival HSC Use 
Case~\#1). Figure~\ref{fig-Brown-astrometry} shows position comparisons in the X-direction (using the 
10265-01-ACS/WFC-F606W  HLA image as the reference) between values from the High Level Science 
Products (HLSP) catalog provided by Brown et~al., and the measurements from the HSC. There are several
differences in these treatments, perhaps the most basic being that Brown et~al.\ combined the 30 different visits 
for each filter into a single deep mosaic image, while the HSC makes 30 separate source lists and then combines 
the results, as described in \S4.1.1.

The upper panel of Figure~\ref{fig-Brown-astrometry} shows the 
resulting comparison for these objects (mainly stars) as a function
of magnitude.  Breaks in magnitude are employed to make it easier
to see how the resulting uncertainties increase for fainter objects,
as expected. However, note that there is also a small fraction of
objects with unexpectedly large errors. Cutout images show that
most of these cases are due to stars with close companions.

The bottom panel shows how the values of the RMS scatter for
individual objects grow from about 2~or 3~mas for
the brightest objects to about 8~mas for the faintest objects for
this particular dataset (i.e., long exposures using ACS/WFC).

\begin{figure}[t]
\centering
\includegraphics[type=\plotext,ext=.\plotext,read=.\plotext,width=\linewidth]{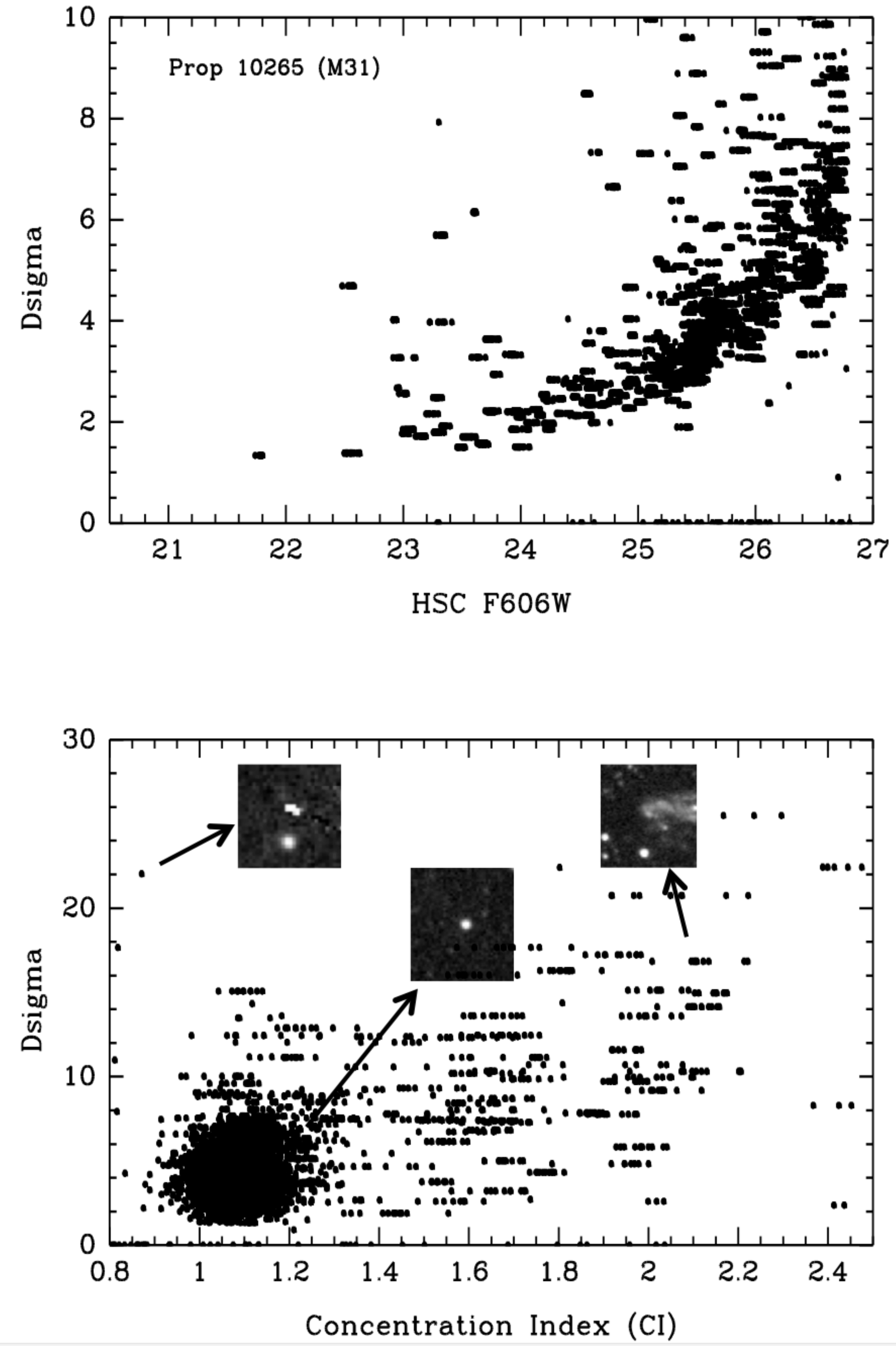}
\caption{Comparisons between Dsigma (in mas), ACS-F606W magnitudes from the HSC (top), 
and Concentration Index (bottom) in the Brown et~al.\ (2009) M31 disk field. See text for description.
\label{fig-Brown-astrometry2}}
\end{figure}

We next turn to other measurements in the HSC for the M31 dataset. The upper
panel in Figure~\ref{fig-Brown-astrometry2} shows how values of Dsigma from
the HSC vary with magnitude.  Dsigma is defined as the RMS scatter
in the individual measurements for each visit (i.e., D) after all
the images have been matched in position (i.e., step~5 discussed
in \S3). Note that the resulting values of Dsigma are similar to
the values of the RMS-X comparison with the Brown et~al.\ (2009)
positions from Figure~\ref{fig-Brown-astrometry}, with median values ranging from about 
2~to 6~mas as a function of magnitude. The bottom panel shows
how the values of Dsigma increase with concentration index (CI),
as expected since the objects with large values of CI are generally
galaxies (as shown in the cutout for one object) or blended stars
where the centroiding is less precise. The vast majority of objects
are isolated stars with values of CI around 1.1 and Dsigma between
2~and 8~mas. A few rare artifacts are also found with discrepant
values such as the detector defect shown in the cutout in the upper left of the right figure.

A general conclusion based on both Figure~\ref{fig-Brown-astrometry}
and~\ref{fig-Brown-astrometry2} is that the limiting
values for the astrometric precisions for a single well exposed ACS
or WFC3 observations in the HSC is a few mas. This agrees with
results we will find using full database comparisons in \S4.3.2.
Astrometric positions for WFPC2 are considerably more uncertain due
to a variety of considerations including larger pixels and more
uncertain geometric solutions, especially for objects that fall on
both the PC and WFC chips in subsequent exposures. This topic will
be discussed in more detail in \S4.3.

\begin{figure}[t]
\centering
\includegraphics[type=\plotext,ext=.\plotext,read=.\plotext,width=\linewidth]{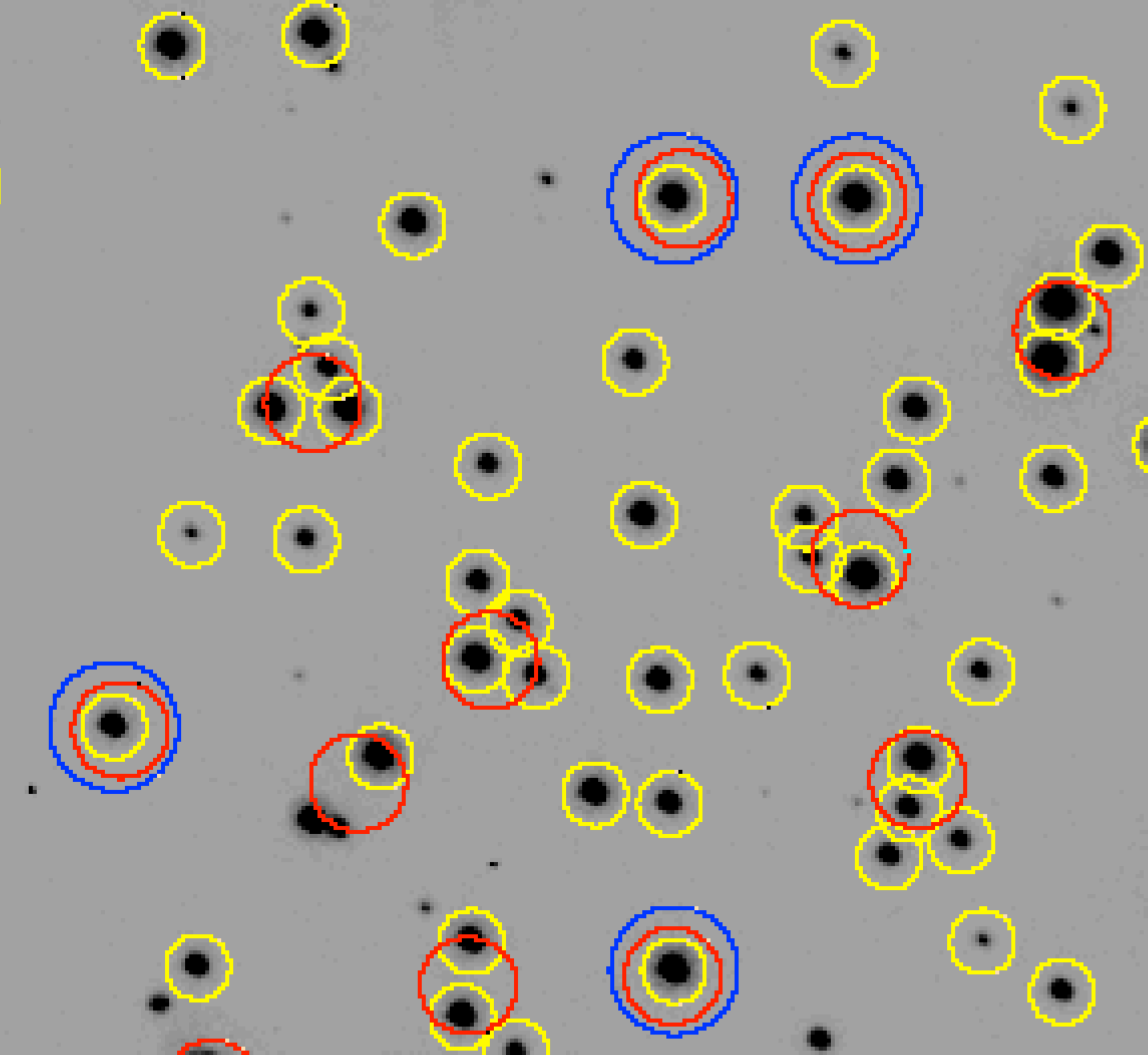}
\caption{Comparison of HSC (yellow) and PanSTARRS (red) astrometry in M4. Blue circles 
show an example of the matches used to measure a mean relative astrometric shift of 
approximately 9~mas between the HSC and PanSTARRS for this field, after filtering out 
poor matches due to differences in spatial resolution. See text for details.
\label{fig-PS-astrometry}}
\end{figure}

\subsubsection{Absolute Astrometry for Point Sources}

As discussed in \S3.1 and shown in Figure~\ref{fig-coverage}, three
different datasets have been used to provide the astrometric backbone
for the HSC, with PanSTARRS being used in the majority of the cases.
Hence, we expect the typical accuracy for the absolute astrometry
of the HSC to be roughly the same as for PanSTARRS. PanSTARRS uses
the 2MASS catalog as its astrometric backbone, and so should have
roughly the same 0.1~arcsec (i.e., 100~mas) absolute astronomical
accuracy (Skutskie et~al.\ 2006). In this section we perform
independent spot checks to make sure the HSC fields are well aligned
with the PanSTARRS sources.

\begin{deluxetable*}{lccccc}
\tablecolumns{6}
\tablewidth{0pt}
\tablecaption{HSC Astrometry Tests Using Radio Catalogs}
\tablehead{
\colhead{Radio Catalog} &
\colhead{\# matches\tablenotemark{a}} & \colhead{RA offset\tablenotemark{b}} & \colhead{Dec offset\tablenotemark{b}} & \colhead{Uncertainty\tablenotemark{c}} \\
\colhead{} &
\colhead{} & \colhead{(arcsec)} & \colhead{(arcsec)} & \colhead{(arcsec)} \\
\colhead{(1)} &
\colhead{(2)} & \colhead{(3)} & \colhead{(4)} & \colhead{(5)} 
}
\startdata
FIRST         & 939 &    $-0.002$ & \phs$0.017$ & $\pm 0.010$ \\
VLA COSMOS    & 469 & \phs$0.030$ &    $-0.074$ & $\pm 0.005$ \\
ICRF2         & 185 & \phs$0.003$ & \phs$0.013$ & $\pm 0.001$ 
\enddata
\tablenotetext{a}{Number of sources matched in the HSC excluding ambiguous matches as
described in the text.}
\tablenotetext{b}{Mean position difference (HSC match position--radio position).}
\tablenotetext{c}{Uncertainty in the mean; i.e., the RMS scatter divided by the square root of the number of matches.}
\label{table-radio}
\end{deluxetable*}

Figure~\ref{fig-PS-astrometry} (from the M4 field discussed in \S4.1.3)
shows that matching between high precision \textit{HST} observations and
ground-based observations can be challenging, especially in crowded
regions. In this figure the yellow circles show the HSC objects and
the red circles show the PanSTARRS objects. In this particular field
more than half of the PanSTARRS objects are clearly blends of several stars when
observed with \textit{HST} resolution. However, by restricting the matches
to have precisions better than 100 mas, and photometrically similar
measurements, we are able to determine good matches for relatively
isolated objects (e.g., the blue circles). Using the whole M4 field
rather than just this small cutout we find relative accuracies of
about 54 mas for single objects, and mean absolute offsets for the ensemble
of HSC and PanSTARRS matches in this field of about 9~mas. Hence,
the agreement between the HSC and PanSTARRS positions is about a
factor of 10 times better than the absolute astrometric accuracy
for PanSTARRS (i.e., 100 mas), and hence does not result in much
additional degradation to the absolute astrometry for the HSC.

Making the same comparison for a variety of other fields (e.g.,
sparser and more crowded fields; galaxies with crowding
or  high background, faint galaxy fields such as the HDF) results
in mean offsets between HSC and PanSTARRS positions with 
values in the range 5~mas (e.g., M87, with isolated, high S/N 
globular cluster) to 15~mas (M83, with crowding and high 
background). We conclude that the absolute accuracy for the 
HSC is essentially the same as for PanSTARRS and 2MASS 
(i.e.,  $\sim$0.1 arcsec).

As discussed in \S3.1, not all HSC fields can be matched with
PanSTARRS, SDSS, or 2MASS. In most cases this is because the
PanSTARRS and SDSS surveys do not cover the particular region
of the sky, or because the density of 2MASS sources is too low to
provide enough matches to make a useful comparison (which is common
for Galactic latitude $|b| \gtrsim 20\arcdeg$). Some \textit{HST} 
observations have such a large mismatch in wavelength, or have such low
quantum efficiency (e.g., far UV observations with WFPC2) that no good 
matches can be found, especially with the near-IR observations from 2MASS. 
These cases are defined with the \texttt{AbsCorr}~= N~flag, and have values 
for absolute astrometry accuracies at the level provided by the HLA (see the 
HLA FAQ discussion). In many cases, such as  early WFPC2 observations, 
this may be 1~or 2~arc seconds. In a few very rare cases absolute
errors up to 10~or 20~arcsec are present,  will be discussed  in \S4.3.3.

\subsubsection{Independent Absolute Astrometric Check using Radio Observations}

The above comparison demonstrates that the HSC is well-aligned to the PanSTARRS 
coordinate system, as expected since most HSC fields in the sky area covered by PanSTARRS 
used that catalog to correct the astrometry.  As a completely independent astrometric test,
we have also matched the HSC to several different radio catalogs. The astrometric calibration 
of radio positions relies on a grid of calibration sources.  The current International Celestial 
Reference Frame (ICRF2; Fey, Gordon \& Jacobs 2009) has an internal accuracy of better 
than 1~mas.  

We cross-matched three different radio catalogs with the HSC.  These catalogs were chosen 
to provide a broad range of tests of the \textit{HST} astrometry:
\begin{itemize}
\itemsep0em
\item The VLA FIRST survey (Becker et~al.\ 1995; White et~al.\ 1997; Helfand et~al.\ 2015)
covers 10,000 deg$^2$ of the northern sky with a FWHM resolution of 5\farcs4 and a source
density of $\sim90$ sources deg$^{-2}$.  Its major advantage is that it covers a wide sky area 
(making it more sensitive to large scale systematics), while its disadvantage is a modest resolution 
that results in RMS accuracies in the radio source positions ranging from 0.1~to 0.5~arcsec 
depending on brightness.

\item The VLA COSMOS survey (Schinnerer et~al.\ 2007) covers the 2 deg$^2$ COSMOS field 
with very deep observations (10 times fainter than FIRST) at a FWHM resolution of 1\farcs8.
Its advantage is that it provides a dense radio catalog (1800 sources deg$^{-2}$) over a region that
also is completely covered by \textit{HST} observations so that there are many HSC-radio matches.
Note that it samples only one spot in the sky, however, so it might not be representative
of the rest of the HSC.

\item The ICRF2 catalog (Fey, Gordon \& Jacobs 2009), as mentioned above, is the basic astrometric 
reference catalog defining the radio coordinate system.  It includes 3414 sources spread over the entire 
sky with extremely accurate radio positions having errors typically less than 1~mas. 
\end{itemize}

\noindent For each radio catalog, all
HSC sources within 8~arcsec of a radio position were first extracted.
The large matching radius was used both to allow for the possibility
of a large difference in the HSC and radio positions, and to identify
cases where the HSC catalog was too crowded to allow a confident
identification of the counterpart. The closest match was accepted
as the optical counterpart of the radio source if it was close enough 
to make the Poisson probability of a chance match $P_F$ less
than 0.03:
 \begin{equation}
P_F(r,N) = 1 - \exp\left[ -N (r/R)^2\right] < 0.03 \quad ,
\end{equation}
where $r$ is the distance to the closest match, and $N$ is the
number of matches within search radius $R=8\arcsec$.  

Most of the rejected objects are radio sources in the nuclei of
galaxies that are resolved by \textit{HST} into a large number of sources.
Clearly many of these are real associations and could be used when 
studying the astrophysics of radio sources, but for the purpose of testing 
the astrometric accuracy of the HSC they can be dropped.  After this 
cut to eliminate ambiguous matches, the remaining contamination by false 
(random) matches ranges from 0.2\% for the ICRF2 to 0.8\% for 
COSMOS to 1\% for FIRST.  

Table~\ref{table-radio} summarizes the results from these three
radio cross-matches.  The mean shifts in the wide area FIRST survey are 
less than 20~mas and are consistent with zero.  The COSMOS field does 
show a significant offset of $\sim80$~mas between the HSC and radio positions.  
That is probably representative of the absolute astrometric accuracy for
a small region of the HSC. The last column  in the table is the uncertainty in the 
mean position.

\begin{figure}[t]
\centering
\includegraphics[type=\plotext,ext=.\plotext,read=.\plotext,width=\linewidth]{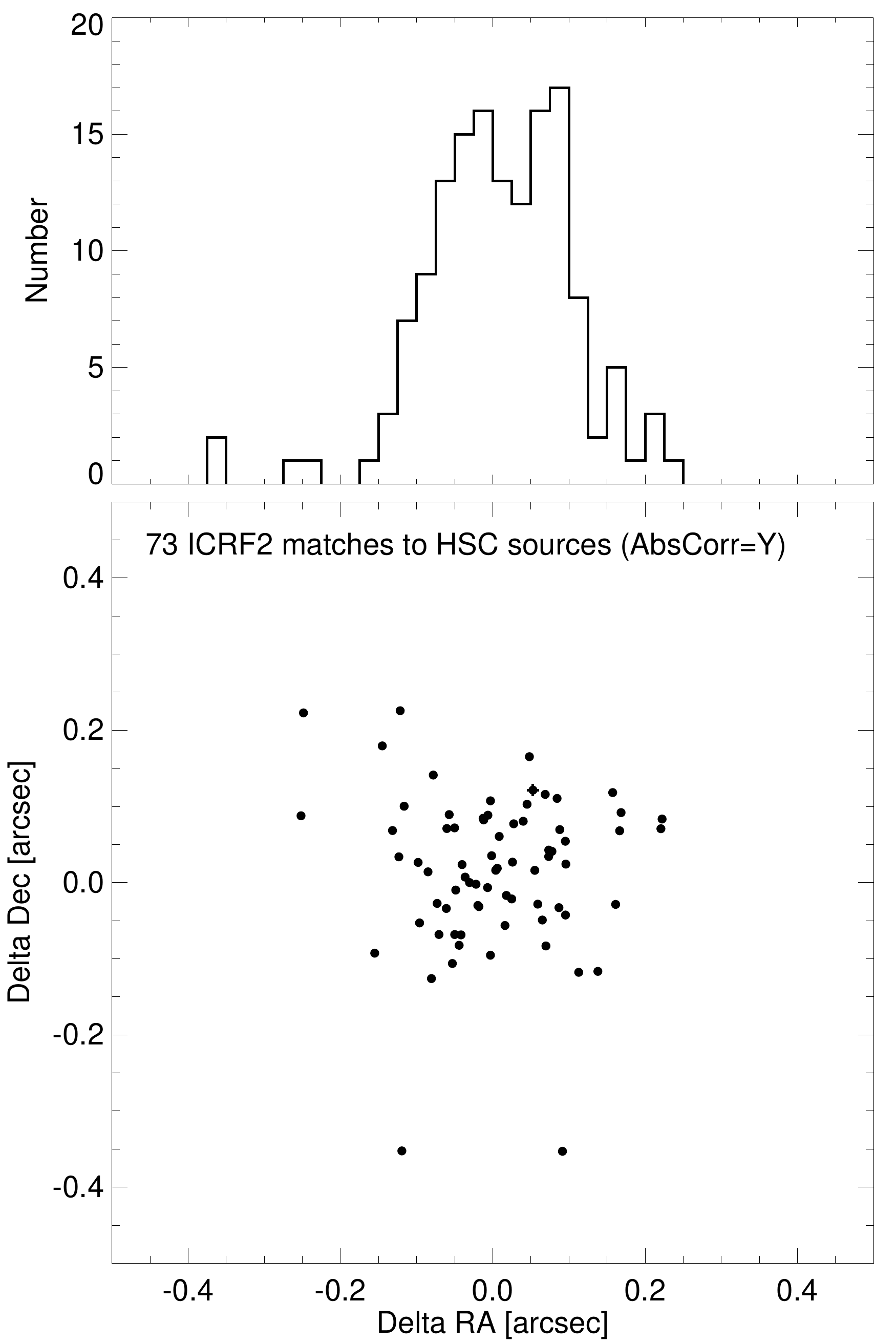}
\caption{Astrometric comparison of the HSC with the ICRF2 radio astrometric reference catalog
(Fey, Gordon \& Jacobs 2009). Only HSC objects with corrected astrometry in unconfused
regions are included (see text for details).  The error bars on the radio positions are shown but
are mostly smaller than the symbols.  The histogram shows the combined distribution of the
one-dimensional offsets in RA and Declination.
\label{fig-radio-astrometry}}
\end{figure}

The ICRF2 catalog shows very small mean offsets compared with the
HSC.  There is however scatter in the positions that is much larger
than the uncertainties in the radio positions (Fig.~\ref{fig-radio-astrometry}).  
The RMS scatter is $\sim0.1$~arcsec. This could be naively interpreted as being 
the corresponding uncertainty in the HSC absolute astrometry, and it does in fact
represent an upper limit on that uncertainty.  However, the reality is more 
complicated.  While the ICRF2 radio sources are very compact objects with 
positions accurately determined from Very Long Baseline Interferometry, 
there is no guarantee that the position of the corresponding optical source 
must match exactly that radio position. The corresponding optical
emission is often not coming from the same physical region but may
instead be from an associated accretion disk, a dense cluster of
stars, or from interstellar gas that is heated and ionized by the
energetic source that powers the radio emission.  Dust in the galaxy
may obscure the nuclear source in the optical and shift its apparent
position.  Consequently, there is ``astrophysical noise'' in these
positions: perfect measurements of any individual radio source
position and its optical counterpart may disagree.  This is an extra
source of scatter in Figure~\ref{fig-radio-astrometry}.

The other source of positional scatter for these ICRF2 sources is
that some of them are very bright, making the HSC positions uncertain. 
Visual examination of the 10~objects with the largest separations in 
Figure~\ref{fig-radio-astrometry} reveals that five are extended galaxies 
(two with dusty disks in the center), two are very bright objects (saturated), 
two are moderately bright (near saturation), and one has no issues and should 
have an accurate HSC position.

In summary, matches to external radio catalogs confirm that systematic astrometric
errors in the HSC are at most 0.1~arcsec, with significant contributions to the 
measured scatter from the radio-optical morphological differences (i.e., ``astrophysical 
noise"). There is no evidence for significant mean offsets over thousands of square degrees
using the FIRST survey.  There is an offset of about 80~mas in comparisons with
the COSMOS deep VLA survey; that is likely to be typical of absolute astrometric 
errors in small regions of the HSC.

\subsection{Photometric and Astrometric Database Comparisons}

Another approach to characterizing the quality of the HSC is to make comparisons using
 measurements from the entire database, rather than the detailed comparisons discussed 
in Sections~4.1 and~4.2. These have the advantage of including a much larger fraction
of the entire database. Hence the approaches are complementary.

\begin{figure}[t]
\centering
\includegraphics[type=\plotext,ext=.\plotext,read=.\plotext,width=0.9\linewidth]{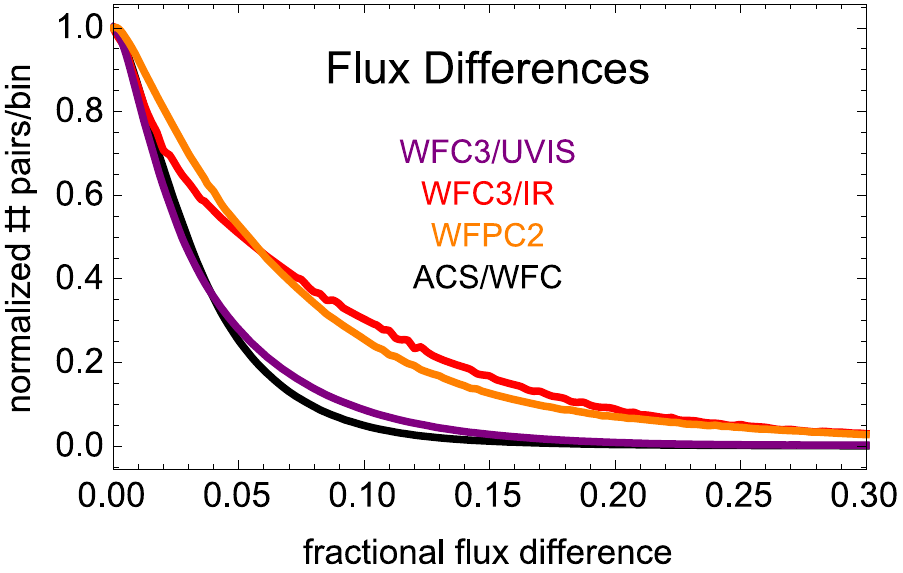}
\caption{Photometric accuracy for Version~1 of the HSC based on repeat measurements using 
the entire database.
\label{fig-photometry-repeats}}
\end{figure}

\subsubsection{Photometric Database Comparisons}

Figure~\ref{fig-photometry-repeats} shows the Version~1 HSC photometric accuracy based on 
repeat measurements for the entire database.  

The data are separated into different detectors and comparisons are made between pairs of  flux estimates 
measured in the same match (i.e., for the same astronomical object) using the large aperture (i.e., 
MagAper2) for the same filter. Only stellar sources (based on measurements of the concentration index)
are used in this comparison,  The x-axis is the flux difference ratio defined as abs(flux1-flux2)/max(flux1,flux2).
The y-axis is the number of pairs of sources per bin normalized to unity at a flux difference of zero.

\subsubsection{Relative Astrometric Database Comparisons}
\begin{figure}[t]
\centering
\includegraphics[type=\plotext,ext=.\plotext,read=.\plotext,width=\linewidth]{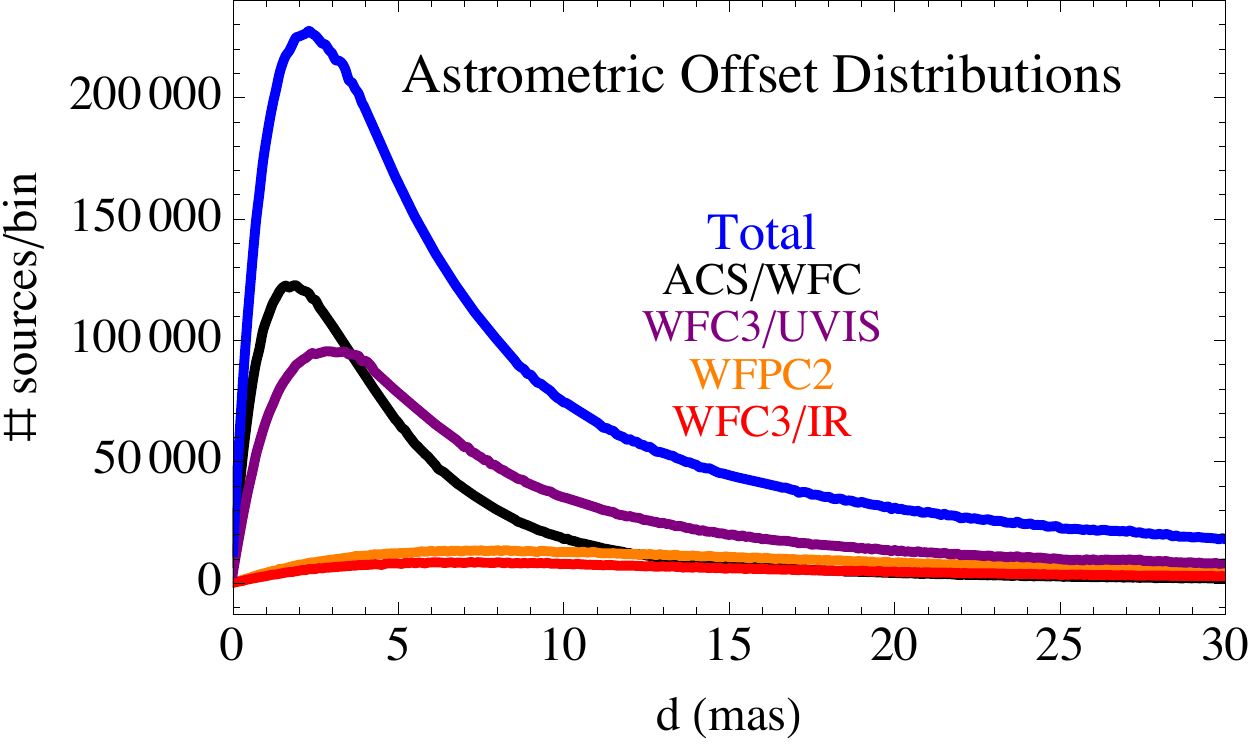}
\caption{Relative astrometric accuracy for Version 1 of the HSC based on repeat measurements using 
the entire database.
\label{fig-astrometry-repeats}}
\end{figure}

Figure~\ref{fig-astrometry-repeats} shows a similar comparison for the entire HSC database for the relative 
astrometry based on repeat measurements, using the white-light detection images. The mode (peak) of the 
distributions for ACS and WFC3 are roughly 2~mas. The upper curve is the same as the HSC corrected curve 
in Figure~\ref{fig-astrometry-offsets}.  The peak of the distributions for the WFPC2 and WFC3/IR occur at 
higher values primarily due to the larger pixels (and lower resolutions) for these instruments.

\begin{figure}[t]
\centering
\includegraphics[type=\plotext,ext=.\plotext,read=.\plotext,width=\linewidth]{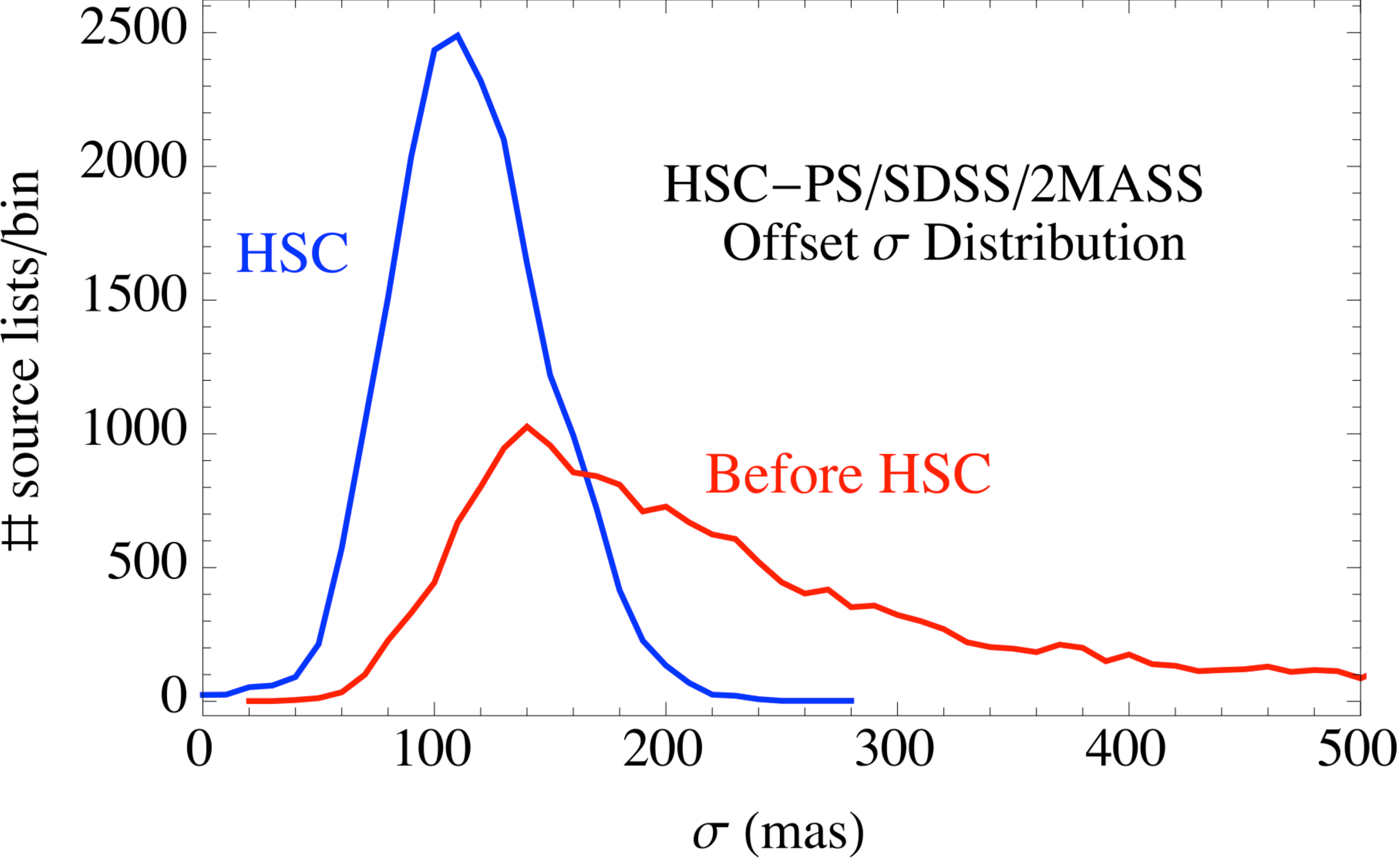}
\caption{Absolute astrometric accuracy for Version~1 of the HSC based on comparisons with 
Pan-STARRS, SDSS, or 2MASS. The red line shows the original distribution from the HLA 
source lists while the blue line shows the distribution after the HSC matching steps have been performed.
\label{fig-astrometry-abs}}
\end{figure}

\subsubsection{Absolute Astrometric Database Comparisons}
As discussed in \S3.1, Step~7 of the HSC pipeline carries out corrections to the absolute astrometry by 
cross matching with external catalogs (PanSTARRS, SDSS, and 2MASS) where possible. We describe 
here the accuracy of the absolute astrometry that has been achieved for the HSC relative to these catalogs. 

The blue line in Figure~\ref{fig-astrometry-abs} shows the resulting absolute astrometry after the HSC 
matching steps described in \S3.1 have been performed. To determine the positional residuals we 
determine the closest external catalog source within 0.3~arcsec of each HSC match position, using the 
same external catalog that was used in Step~4 in \S3.1 for that source list. The resulting distribution is 
quite tight, with mean and median values near 120~mas, in general agreement with the results from 
\S4.2.3 based on comparisons with radio observations. The standard deviation is just 34~mas.

The red line in Figure~\ref{fig-astrometry-abs} shows the absolute astrometry before the HSC matching 
steps described in \S3.1 have been performed. The resulting distribution is much broader, with a mean 
value of 450~mas and median value of 220~mas. The large difference between the mean and median values 
show that there are some very large residuals, in a few rare cases more than 10~arcsec. This is also reflected 
in the standard deviation, which is 1.1~arcsec for the uncorrected HLA positions. 

\begin{figure}[t]
\centering
\includegraphics[type=\plotext,ext=.\plotext,read=.\plotext,width=\linewidth]{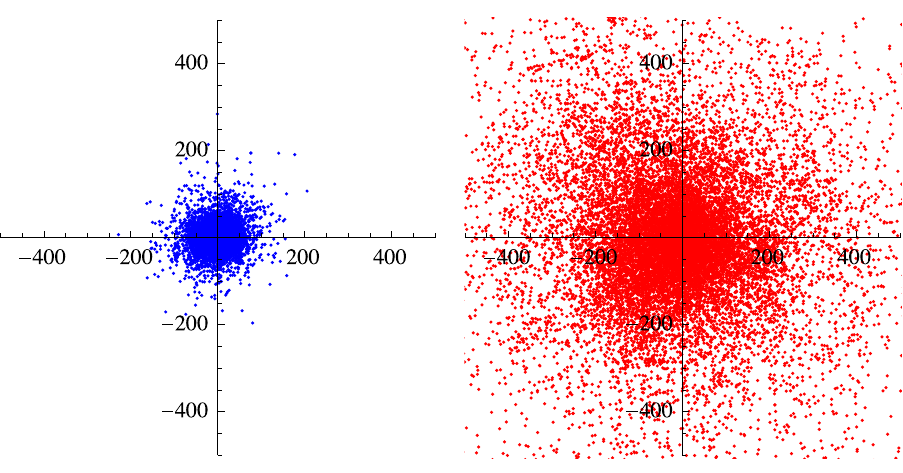}
\caption{A plot of the mean offsets in source positions with respect to the  PanSTARRS, SDSS, and 2MASS 
catalogs used to determine the match (see \S3.1). The horizontal axis is right ascension in mas while the vertical 
axis is declination in mas. The blue points on the left are for the HSC (i.e., after corrections have been made) 
while the red points on the right are for the HLA positions. 
\label{fig-astrometry-jitter}}
\end{figure}

Figure~\ref{fig-astrometry-jitter}  plots the mean offsets in mas for the right ascension (horizontal axis) and 
declination (vertical axis). Each plotted point corresponds to a single white-light source list. The left panel 
(blue) shows the offsets after HSC corrections and the right panel (red) is before HSC corrections. The right 
panel has a halo that extends well beyond the plotted region. The results show that the uncorrected
astrometric offsets have a much broader central core than the corrected ones and also have a long tail.
The results also suggest that a mean overall shift is present in the uncorrected \textit{HST} astrometry that
 is not present after correction. Among source lists with positional shifts less than 300~mas, the mean (RA, Dec) 
shift is ($-$11,$-$7)~mas before correction and (0.01, 0.7)~mas after correction.

We conclude that the HSC has typical internal photometric accuracies better than 0.1~mag, relative astrometric 
accuracies of $\sim$10~mas, and absolute astrometric accuracies of $\sim$100~mas, although in specific regions 
the accuracies can be better or worse, due to the inherent non-uniformity of the HSC.  These values are in good
agreement with the spot checks shown in \S4.


\subsection{Incompleteness}

The HSC is incomplete for a number of reasons. For example, only three of the 12~instruments flown on 
\textit{Hubble} are included; WFPC2, ACS/WFC, and WFC3. However, these are the three instruments with
the largest numbers of \textit{Hubble} detections, hence contain the majority of all the sources ever observed by 
\textit{HST}. Future plans call for the inclusion of NICMOS and ACS/HRC observations, and possibly others
in the future (e.g., STIS imaging).

It is also important to remember that even for the three instruments included in Version~1, only about 65\% of the
 ACS/WFC, WFPC2, WFC3 images are included in the catalog due to image quality and other issues (see \S3.1 
and 4.2.3 for a discussion). In addition, as will be stressed in Section~5, the quality and depth of the source lists for 
the three instruments is non-uniform. While this will be improved in the future, the HSC will always have different
 completeness thresholds in different regions for a number of reasons, including the different quantum efficiencies 
of the instruments (i.e., WFPC2 is much shallower than ACS and WFC3) and the wide range in exposure times.

For these and other reasons, researchers should be aware that just because a source is not in version~1 of the HSC 
does not mean that there is no  \textit{Hubble} observations of it. The HLA can be used to make a more complete
search, but for a definitive determination (e.g., when checking for duplications when writing \textit{HST} 
observing proposals), the MAST archive tools must be used.

\section{Caveats and Warnings}\label{sec:science}
\begin{figure*}[t]
\centering
\includegraphics[type=\plotext,ext=.\plotext,read=.\plotext,width=0.9\linewidth]{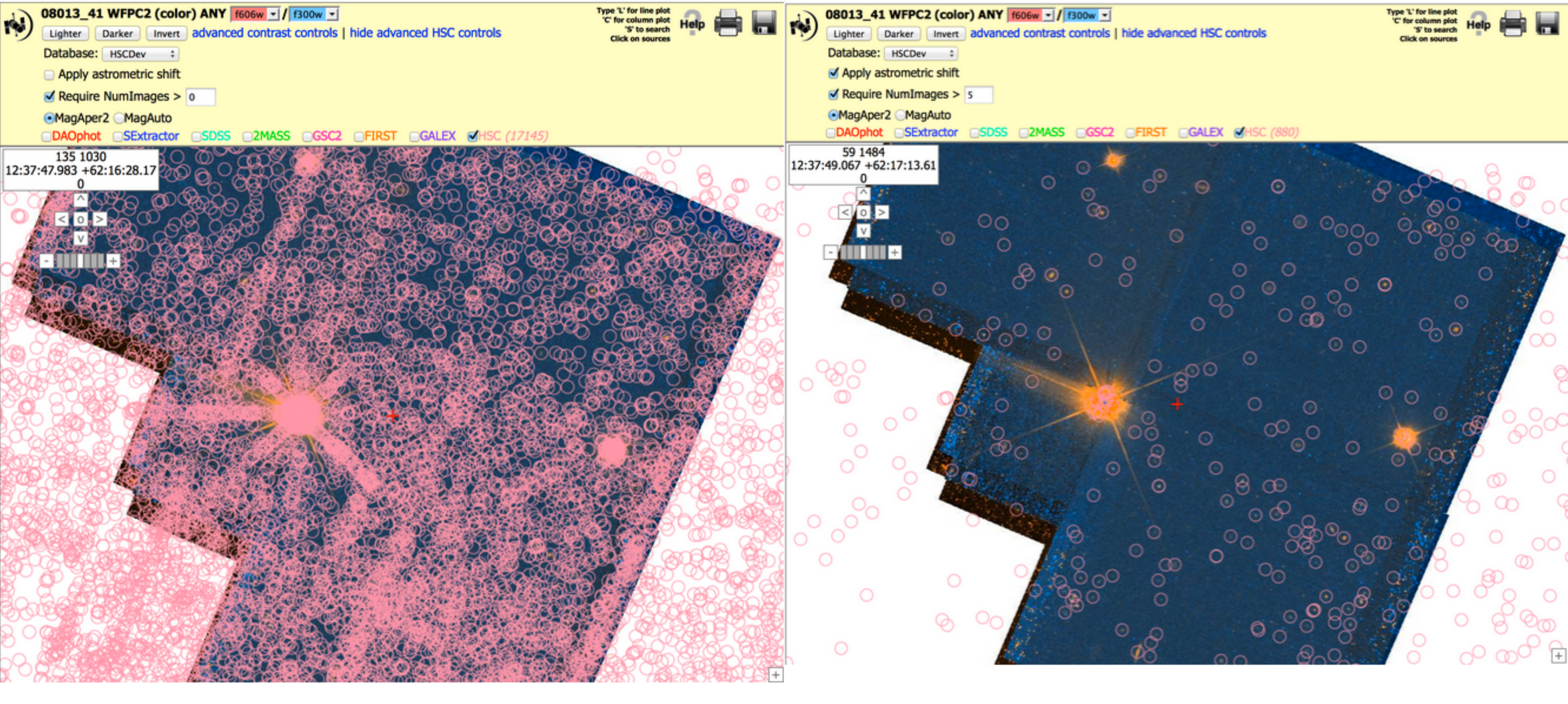}
\caption{Example of a particularly bad WFPC2 source list (left image) showing artifacts from bright stars and edge effects.
The small pink circles are objects in the HSC. Using \texttt{NumImages} $>5$ (right image) removes most of these artifacts.
\label{fig-bad-WFPC2}}
\end{figure*}

\begin{figure*}
\centering
\includegraphics[type=\plotext,ext=.\plotext,read=.\plotext,width=0.9\linewidth]{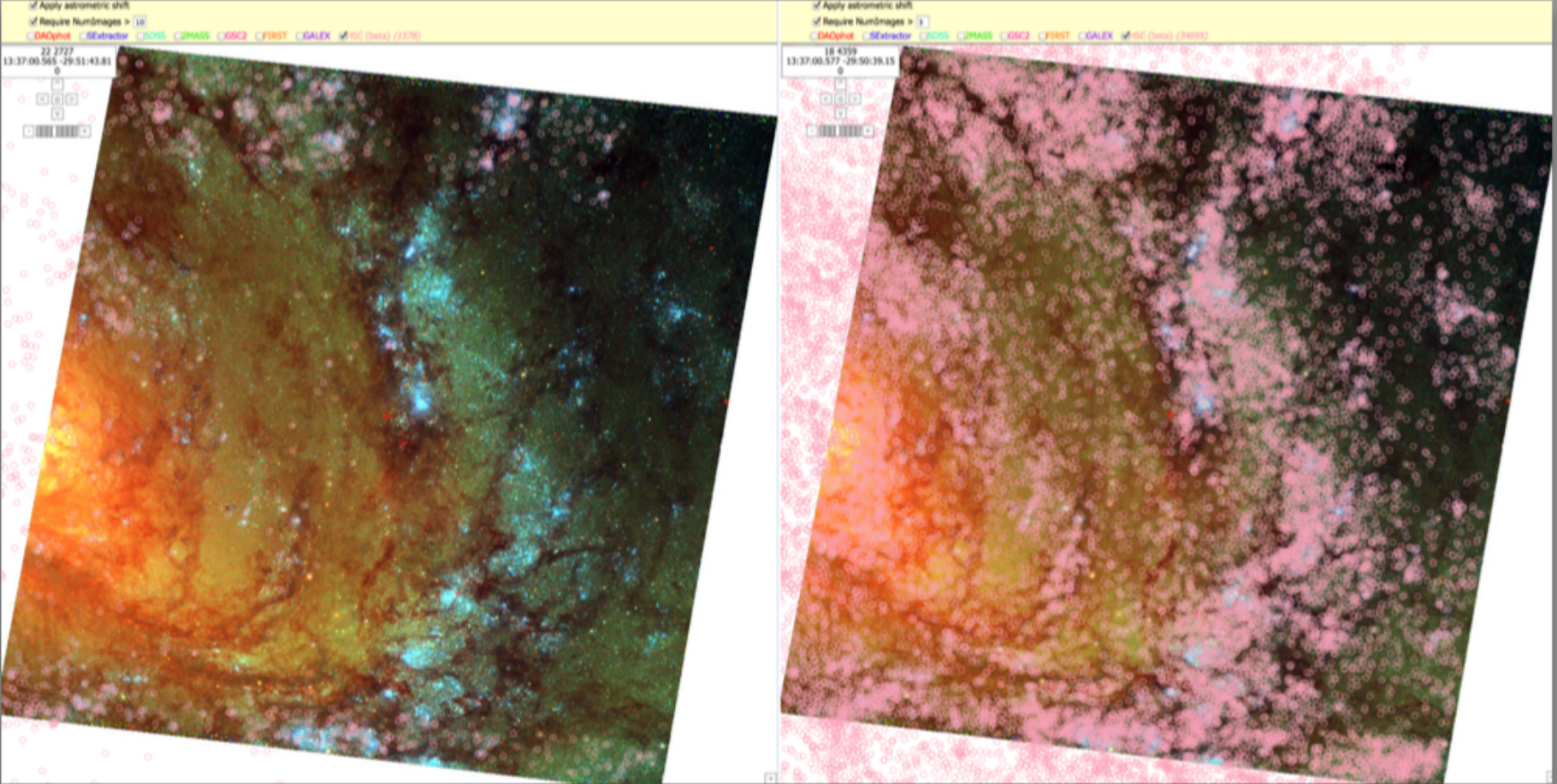}
\caption{An example of the non-uniformities that are possible using improper search criteria, in this case 
\texttt{NumImages}~$>10$ (left image) rather than $>3$ (right image). Additional source lists from overlapping 
HLA images in the upper and lower parts of the galaxy (M83), images not shown here, result in various corners and 
linear features in the left image.
\label{fig-nonuniform}}
\end{figure*}

\begin{figure*}
\centering
\includegraphics[type=\plotext,ext=.\plotext,read=.\plotext,width=0.7\linewidth]{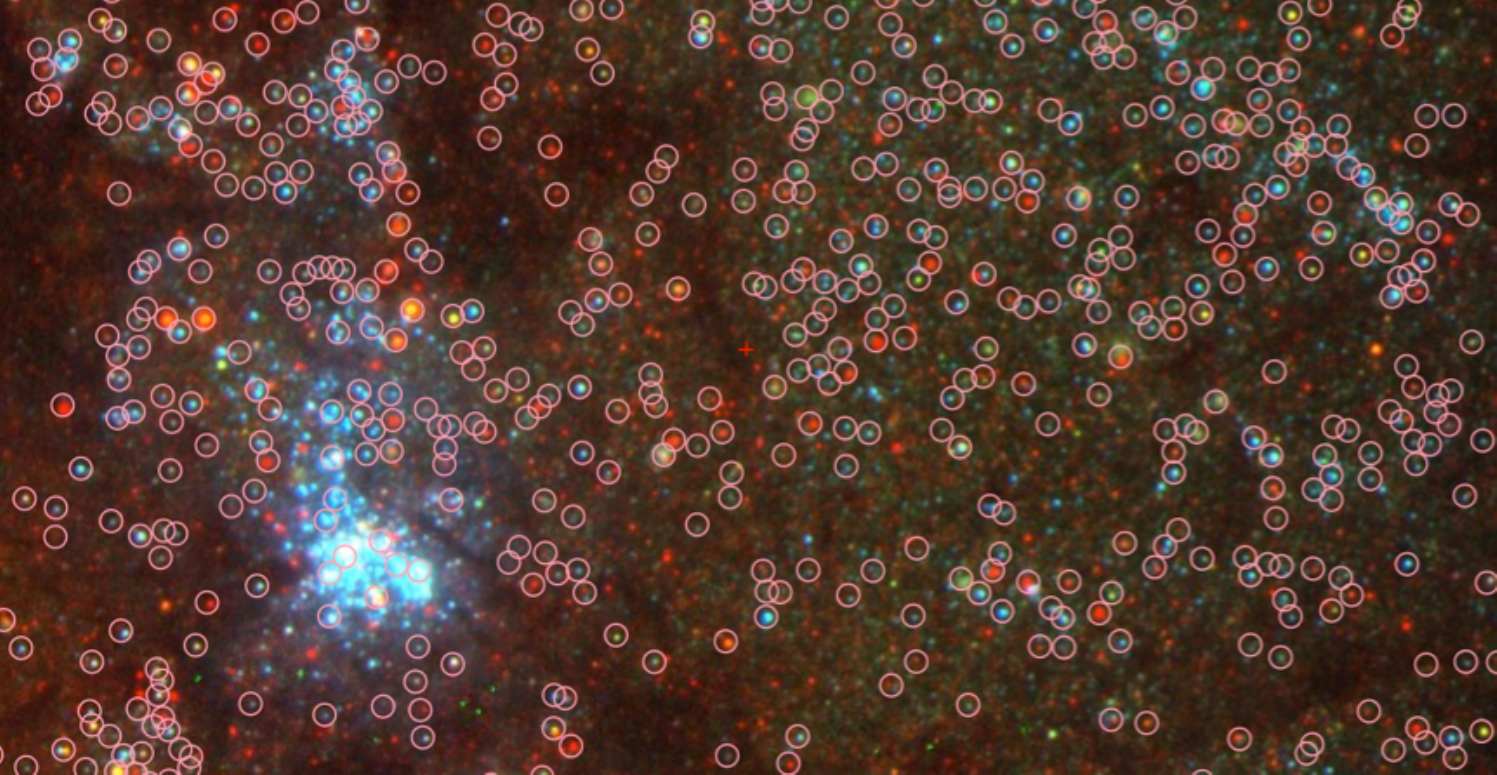}
\caption{A blowup near the center of the right panel in Figure~\ref{fig-nonuniform}. Note that while the catalog is 
quite good in general, it is missing some stars in regions of high background.
\label{fig-nonuniform-blowup}}
\end{figure*}

As stressed in many sections of this paper, the HSC is not a typical wide-area, uniform catalog such as 2MASS, 
SDSS, or PanSTARRS. It is based on a diverse set of observations using pencil-beam exposures covering only a 
small fraction of the sky. While it has tremendous potential for doing science, it can also easily be misused. Users 
should not simply use the HSC as a database search tool. They need to:

\begin{itemize}
\item View the HSC overlaid on images. While the vast majority of the source lists are quite good, there are also 
problem areas that can contain obvious artifacts (e.g., see Figures~\ref{fig-bad-WFPC2}, \ref{fig-nonuniform}, and 
\ref{fig-nonuniform-blowup}).

\item Try different selection filters (e.g., \texttt{NumImages}~$>$ some number) to see how it affects the 
science results. Other potentially useful selection filters are discussed in \S4.1.4 . In many cases results from 
observations using different instruments can also be compared.
\end{itemize}

\begin{figure*}
\centering
\includegraphics[type=\plotext,ext=.\plotext,read=.\plotext,width=0.7\linewidth]{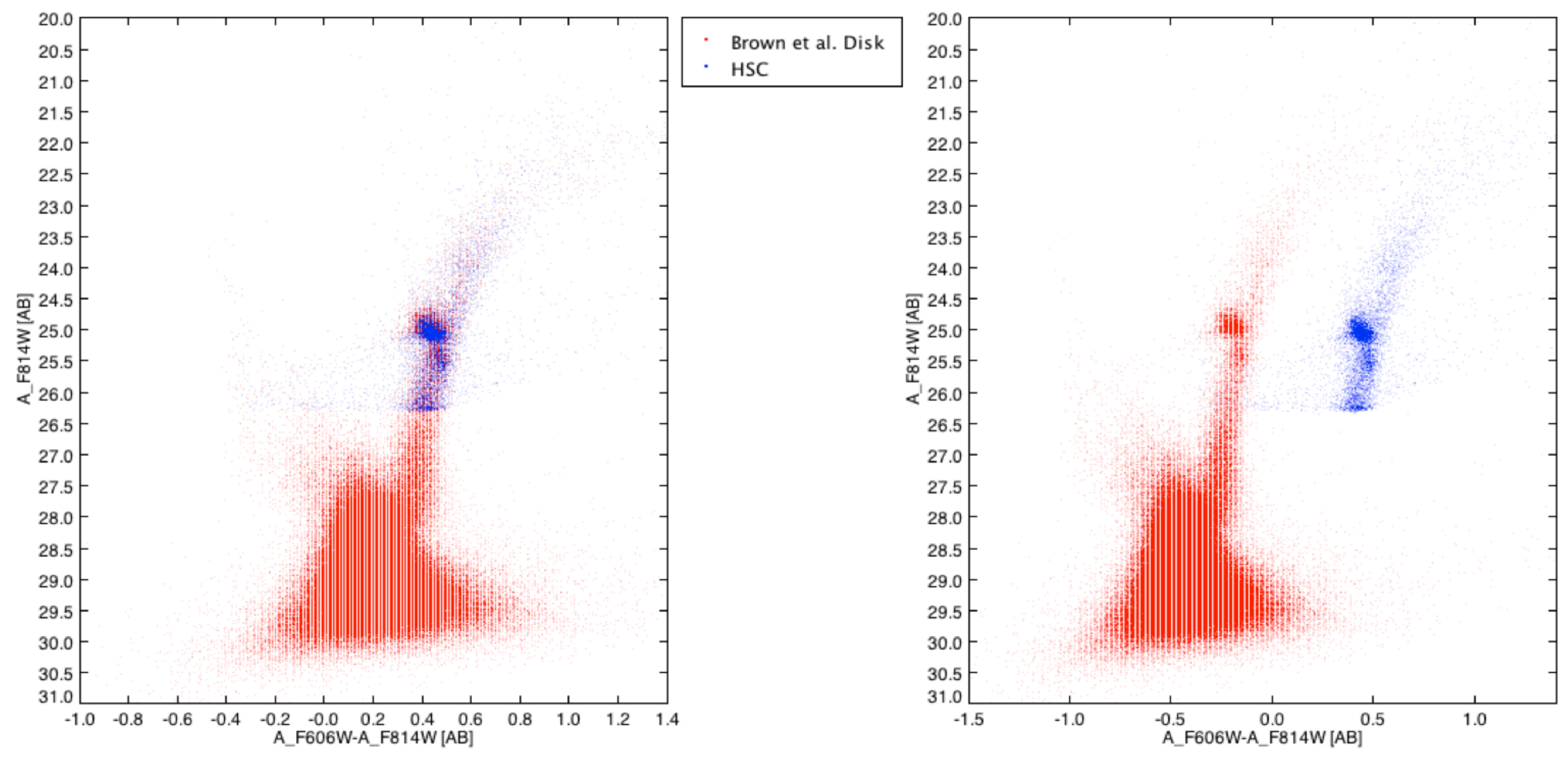}
\caption{Comparison of HSC photometry (blue) and Brown et. al., (2009) photometry (red). An offset  has 
been added in the right panel to make the comparison of small details easier. See Figures~\ref{fig-M31-image} 
and \ref{fig-M31-photometry2}, HSC Use Case~\#1, and HSC Archival Use Case~\#1 for more details. 
\label{fig-Brown-deep-photometry}}
\end{figure*}

\newpage
\subsection{Five Things You Should Know about Version 1 of the HSC}

New users should keep the following in mind when using the HSC.
\begin{enumerate}
\item Detailed use cases and videos are available for training. See Appendix~\ref{sec:appendix_D} for pointers.

\item Coverage in certain regions can be very non-uniform (unlike surveys such as SDSS), since source lists have 
been combined for pointed observations from a wide range of \textit{HST} instruments, filters, and exposure times.

\item WFPC2 and ACS source lists are of poorer quality than WFC3 source lists. As we have gained experience the 
HLA source lists have improved. For example, many of the earlier limitations (e.g., depth, difficulty finding sources in 
regions of high background, edge effects, \dots) have been improved in the WFC3 source lists. These improved 
algorithms will be used when making new WFPC2 and ACS source lists, and will be incorporated into a future release 
of the HSC.

\item The default is to show all HSC objects in the catalog. This may include a large number of artifacts. You can 
request \texttt{NumImages}~$>1$ (or more) to filter out many artifacts in the HSC. (But in regions covered by only a 
single \textit{HST} filter, this will remove all HSC sources.)

\item The default is to use aperture magnitudes (i.e., \texttt{MagAper2}) in the ABMAG system. Transformations 
are necessary to convert to other systems (e.g., VEGAMAG or STMAG), or from one instrument to another, or to
other photometric systems (e.g., Johnson-Cousins or SDSS ugriz). Transformations are available in a variety of 
references (e.g., Holtzman et~al.\ 1995 for WFPC2 and Sirianni et~al.\ 2005 for ACS), but for a generic reference  
stsdas.synphot can be used.  Aperture corrections are needed to convert aperture magnitudes to total magnitudes for 
stars.  For extended sources \texttt{MagAuto} can be requested.
\end{enumerate}

\section{Summary and Future Plans}\label{sec:summary}

Version 1 of the Hubble Source Catalog includes WFPC2, ACS/WFC and WFC3 photometric measurements based 
on SExtractor source lists from data release DR8 of the Hubble Legacy Archive.  The current version of the catalog
includes roughly 80~million detections of 30~million objects involving 112 different detector/filter combinations and 
about 160 thousand \textit{HST} exposures.  The mean photometric accuracy is better than 0.10~mag and the relative 
astrometric residuals are typically within 10~mas. Better precision (e.g., to 0.02~mag and 2~mas or better) is often 
possible in certain circumstances (e.g., for bright isolated stars).  The absolute astrometric accuracy is better than 0.1~arcsec
in most cases.

A number of improvements and enhancements for the HSC are planned for the future. In the relatively short term, the 
primary improvement will be to upgrade the WFPC2 and ACS source lists using the algorithms developed for the WFC3.  
We also plan to incorporate HLA source lists for observations taken with the ACS High Resolution Camera (ACS/HRC) 
and Near Infrared Camera and Multi-Object Spectrometer (NICMOS).

A more fundamental improvement planned for the future is to use the precise offsets determined for the HSC to combine 
the visit-based images into deeper mosaics. HLA source lists will then be obtained using these images to develop much 
deeper catalogs (e.g., see Figure~\ref{fig-Brown-deep-photometry}, where a mosaic image goes roughly four magnitudes 
deeper). ``Forced photometry" at the locations determined from the mosaic images will then be performed for all exposures. 
This procedure will also provide better information about nondetections and upper limits.

Other planned additions are the incorporation of PSF-fitting photometry using the Anderson et~al.\ (2008) photometry 
routines, and integration of spectroscopic information into the HSC.

In the near future, it will be possible to improve the absolute astrometry of Pan-STARRS by linking it to the \textit{Gaia} 
catalog.  Then, by matching the HSC with the improved Pan-STARRS catalog, the resulting absolute astrometric accuracy 
of the HSC will improve from $\sim$0.1 to  $\sim$0.01~arcsec. One of the motivations for improving the absolute astrometry  
is to make more reliable identifications of \textit{HST} sources with objects observed by other telescopes at different wavelengths 
(e.g., X-ray, UV, IR, mm, and radio).  Improving the HSC absolute astrometry will ensure that this term of the error budget is
negligible compared to the absolute astrometric accuracies of most other  telescopes, including \textit{Chandra}, \textit{Spitzer}, 
and \textit{GALEX}, and will even provide accuracies comparable with that of mm and radio interferometers such as ALMA and 
JVLA.

The tools used to access the HSC will also be enhanced in the next few years. One of the primary goals is to better integrate the 
tools discussed in \S5 (the MAST Discovery Portal, the HSC CasJobs service, the HSC home page and the HLA Interactive 
Display) so that a single interface will allow users easy access to most of the capabilities that are currently distributed across four 
separate interfaces. Another challenge on the longer term will be to develop tools to more easily combine and compare 
multiwavelength data sets (e.g., with different spatial resolution) and multi-dimensional data-cubes (e.g., from ALMA and 
\textit{JWST\/}).


We encourage the development of value-added-projects based on the HSC database.  An example that is already under 
development is an ESA-based project to develop a Hubble Catalog of Variables
(\url{http://www.spacetelescope.org/forscientists/announcements/sci150008/}). Other possibilities that are under 
discussion are the determination of transformation equations to support the combination of data from different instruments, 
and determinations of photo-Z redshift estimates based on HSC data.  We expect that in many cases the products of the
value-added projects will be integrated into future version of the HSC.

Catalogs have been a mainstay in astronomy for centuries.  Historical examples include the Messier, Herschel and New 
General Catalogs. More recent examples include 2MASS, \textit{Hipparcos}, and SDSS. In many ways the Hubble Source 
Catalog will be unique, first and foremost because of the depth and spatial resolution of the \textit{Hubble} image. The HSC 
will be an important reference for future telescopes, such as the \textit{James Webb Space Telescope}, and survey programs, 
such as LSST.

In this paper we have attempted to demonstrate the potential of the HSC while also educating HSC users regarding possible 
pitfalls.  The key point is that by its very nature (i.e., deep pencil-beam observations using a wide variety of instruments and 
observing modes), the HSC is a very different database than most other surveys that have uniform ``all-sky'' coverage (e.g., SDSS).  
While the diversity of the HSC dictates the need for caution when developing queries and analyzing data, it also provides the 
opportunity for cross checking the results in many cases.

Astronomers will use the HSC in different ways. At the most basic level it provides a quick way to determine what \textit{Hubble}
observations have been taken of an object. When building your own catalogs, the HSC can be used as a consistency check. Some 
people will use the HSC to do feasibility checks, and to perform preliminary analysis. In other cases users will be able to use the catalog 
to address their primary science goals.

We expect the quality of the HSC to continually improve, as known problems are fixed and new reduction techniques are incorporated.  
While care will be required in using the Hubble Source Catalog, due to its inherent non-uniformity, it is clear that the HSC provides a 
powerful new tool for research with \textit{Hubble} data.

\acknowledgments
We  thank the referee for helpful suggestions that improved the paper. The HSC is based on observations made with the NASA/ESA 
\textit{Hubble Space Telescope}, and obtained from the Hubble Legacy Archive, which is a collaboration between the Space Telescope 
Science Institute (STScI/NASA), the Space Telescope European Coordinating Facility (ST-ECF/ESAC/ESA) and the Canadian 
Astronomy Data Centre (CADC/NRC/CSA).

\appendix

\section{Comparisons with Studies Based on Use Cases}\label{sec:appendix_A}

In this section we examine three specific science projects to see whether using the HSC would give similar results. 

\subsection{Brown et~al.\ (2009)---Color Magnitude Diagram in the Outer Disk of M31}

Figure~\ref{fig-Brown-deep-photometry} shows a comparison of HSC photometry with the Brown et~al.\ (2009) study of an 
outer disk region in the Andromeda galaxy.  At the brighter magnitudes the comparison is quite good, as discussed in \S4.1.1. 
However, the Brown et~al.\ data is deeper than the HSC by approximately four magnitudes, as expected since this comes from a 
mosaic where all 30~visits are co-added together compared to the HSC where the measurements are from individual one-orbit visits.  
However, it should be noted that the photometric uncertainties at these very faint magnitudes are very large, which precludes the use 
of these stars for much more than counting purposes.

More typically, \textit{Hubble} observers employ just one or two longer visits, hence the difference between the HSC and co-added 
mosaics is generally much smaller. Nevertheless, HSC users should be aware that it is often possible to go deeper, or to obtain more 
precise measurements than possible from the general-purpose catalogs produced by the HLA using only the visit-based source lists.

See Figure~\ref{fig-M31-photometry2} for a one-to-one comparison of magnitudes for individual sources, and HSC Use Case~\#1 
and HSC Archival Use Case~\#1 (using the Beta~0.2 version of the HSC) for more detailed discussions.

\subsection{Bernard et~al.\ (2010)---Variability in IC 1613}

\begin{figure}[p]
\centering
\includegraphics[type=\plotext,ext=.\plotext,read=.\plotext,width=0.8\linewidth]{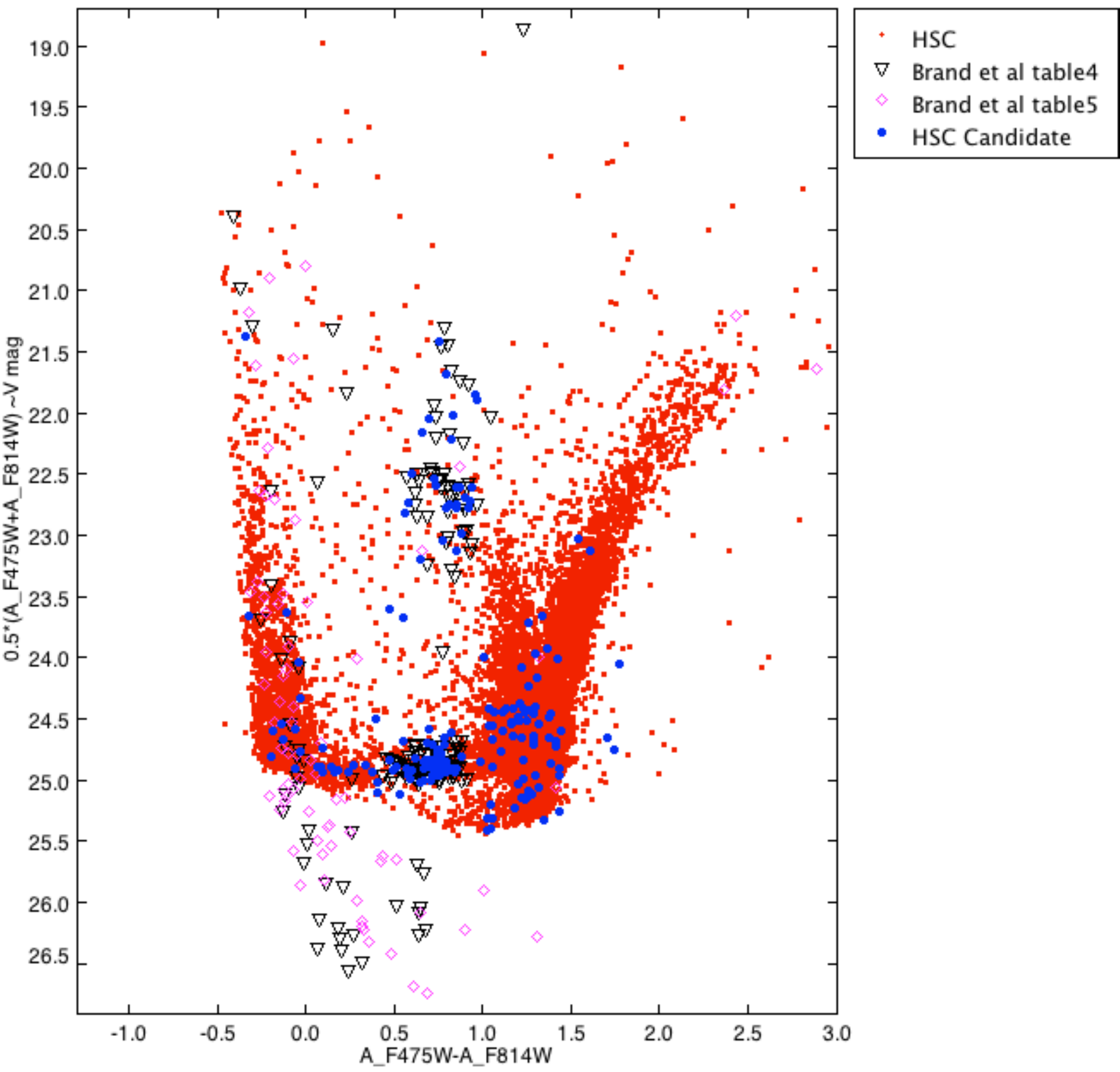}
\caption{The Color-magnitude diagram for IC~1613. HSC non-variable stars are shown in red while HSC candidate variables are 
plotted in blue. Variables stars from Bernard et~al.\ (2010) are included (see insert for symbol definitions). The vertical dashed lines
roughly delimit the instability strip. Note that the HSC candidate variables are found in the same part of the diagram as the Bernard
et~al.\ variables, but do not go as deep. However, the regions of the diagram containing Cepheids and RR~Lyrae variables are well
covered within the HSC limiting magnitude in this region. See HSC Use Case~\#3 and archival HSC Use Case~\#2 for details.
\label{fig-IC1613-CMD}}
\end{figure}

Figure~\ref{fig-IC1613-CMD} shows the result of a search of the HSC for variable stars in the dwarf galaxy IC~1613, a field studied 
in detail by Bernard et~al.\ (2010).  In total we find 210 candidate variable stars from the HSC, compared to 259 found in the Bernard
et~al.\ (2010) paper.  HSC Use Case~\#3 uses this dataset to show a simple version of how to find variable stars using the HSC, while
HSC Archival Use Case~\#2 shows a detailed treatment using the Beta~0.2 HSC database.

\begin{figure*}[p]
\centering
\includegraphics[type=\plotext,ext=.\plotext,read=.\plotext,width=0.9\linewidth]{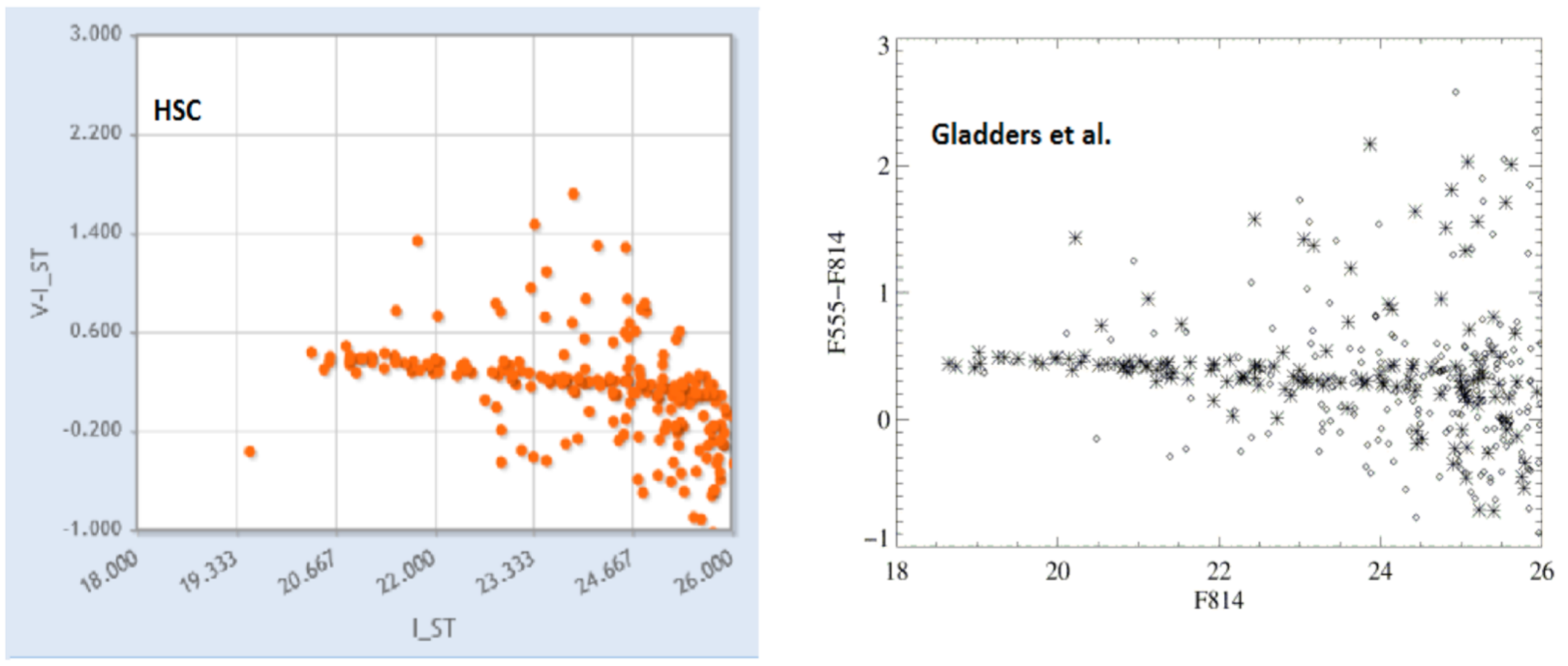}
\caption{The red sequence for galaxies in Abell 2390 from HSC Use Case~\#6 (left), showing good agreement with the original 
Gladders et~al.\ (1998) study (right). The HSC magnitudes are  shifted slightly relative to the Gladders et~al. plot since the HSC 
uses ABMAG and Gladders uses STMAG magnitudes, as discussed in more detail in the use case. 
\label{fig-A2390-red}}
\end{figure*}

\subsection{Gladders et~al.\ (1998)---The Slope of the Elliptical Red Sequence in Abell 2390}

Figure~\ref{fig-A2390-red} shows a re-creation using the HSC of the study of the red sequence for elliptical galaxies in Abell 2390 
by Gladders et~al.\ (1998). After applying aperture corrections from the \textit{HST} exposure time calculator, as provided in the 
HSC FAQ, and extinction values from Schlegel et~al. (1998), as provided in the HSC Summary Form, the agreement with the slope 
found by Gladders et~al.\ is quite good (i.e., $m=-0.042 \pm0.007$ using the HSC and $m=-0.037\pm0.004$ from Gladders et~al.\ 
1998).  See HSC Use Case~\#6 for a more detailed discussion.

\newpage

\section{Tools for Accessing the HSC}\label{sec:appendix_B}

There are four ways to access Version~1 of the HSC. This is partly due to the historical evolution of the tools, but also reflects the
need to provide different types of services, as described below. See Appendix~\ref{sec:appendix_D} for a summary of URLs to 
access the HSC tools and related sites.

\subsection{MAST Discovery Portal (Browsing, Filtering, Plotting, and Cross Matching)}

\begin{figure}[t]
\centering
\includegraphics[type=\plotext,ext=.\plotext,read=.\plotext,width=\linewidth]{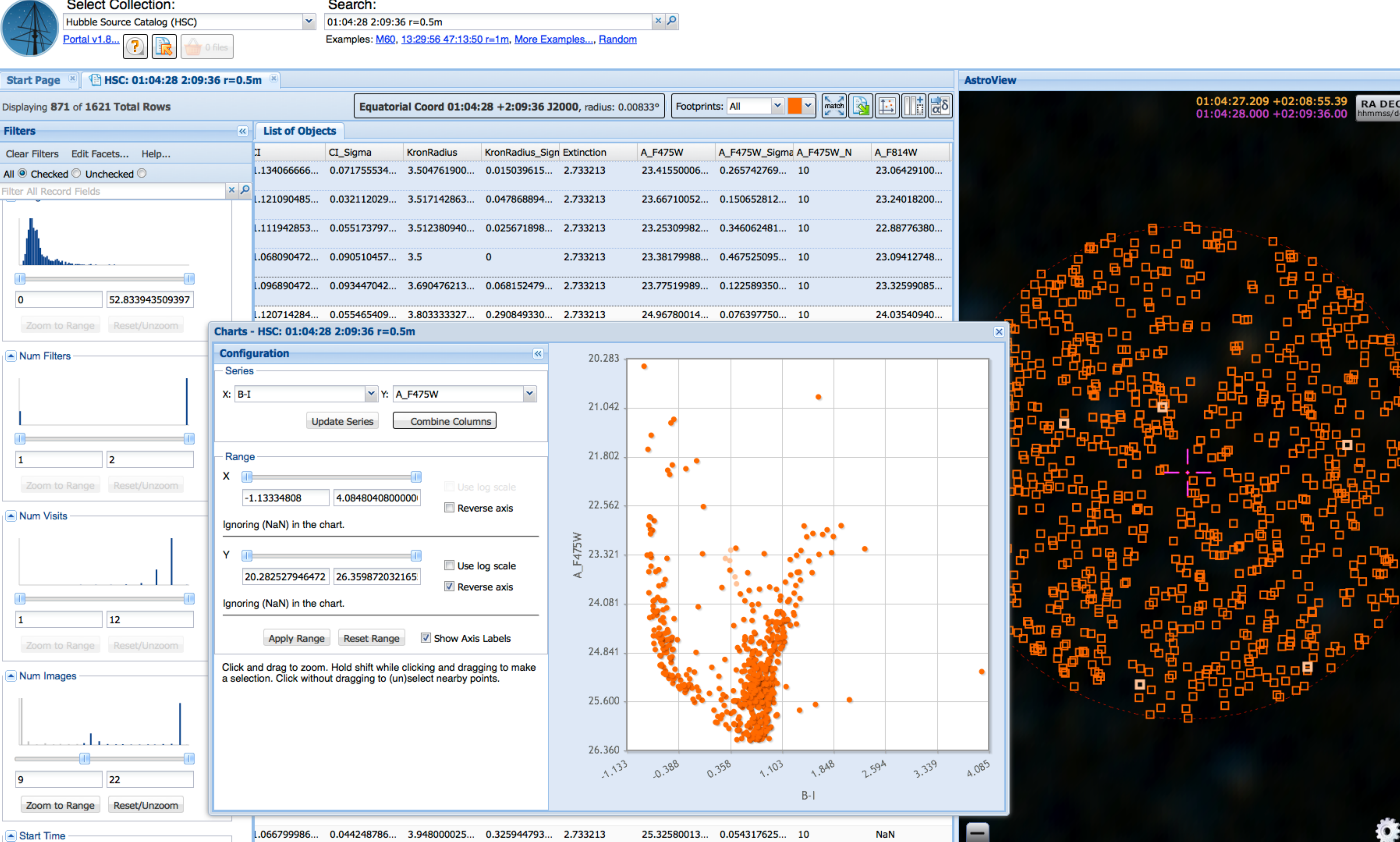}
\caption{Screenshot from HSC Use Case~\#3 showing various aspects of the MAST Discovery Portal.
\label{fig-portal}}
\end{figure}

The primary access tool for the HSC is the MAST Discovery Portal (\url{http://mast.stsci.edu}). This generally provides 
the best way to browse what is in the HSC and to do some quick plotting and/or cross matching with other data. It also 
allows users to download the needed data for further analysis. Its primary current limitation is that only 10,000 sources can 
be included in a given search, and only \texttt{MagAper2} (aperture magnitudes using the larger aperture), rather than 
\texttt{MagAuto} (extended photometry) magnitudes are included. However, as discussed in the next section, CasJobs can 
be used to obtain larger samples, and/or retrieve values of \texttt{MagAuto}. These values can then be filtered and uploaded 
into the Discovery Portal if desired.

Originally developed as part of the Virtual Observatory initiative, the Portal is now the principal access tool for all MAST data. 
It has been modified to include access to HSC data and to include features needed to view \textit{HST} images. It includes a 
wide range of tools for viewing, filtering (e.g., setting a minimum \texttt{NumImages} to remove residual cosmic rays and other 
artifacts), plotting, cross-matching, and downloading.

Figure~\ref{fig-portal} shows an example of how the Discovery Portal (shown as it appeared for the Version 1 release---modifications
can be expected in the future) can be used to find variable stars in IC~1613 (from HSC Use Case~\#3).

\subsection{CasJobs (Advanced Search and Analysis)\label{sec-casjobs}}

The Catalog Archive Server Jobs System (CasJobs) was developed by the Johns Hopkins University/Sloan Digital Sky Survey 
(JHU/SDSS) team. With their permission, MAST has used CasJobs to provide database query access to several MAST databases, 
including \textit{GALEX}, \textit{Kepler}, and most recently the HSC (\url{http://mastweb.stsci.edu/hcasjobs}). The purpose 
of CasJobs is to permit large queries, phrased in the Structured Query Language (SQL), to be run in either real time or in batch 
queues. Therefore, it does not have the limitations of only including a small subsample of the HSC, as is the case for the MAST 
Discovery Portal. CasJobs queries may run for hours and may produce large output tables with millions of sources that are stored 
in the user's MyDB work area. CasJobs is a very powerful interface.  However, it is  more difficult to learn to use than the Portal 
and also does not have the wide variety of graphic tools available in the Discovery Portal.

\begin{figure}[t]
\centering
\includegraphics[type=\plotext,ext=.\plotext,read=.\plotext,width=0.8\linewidth]{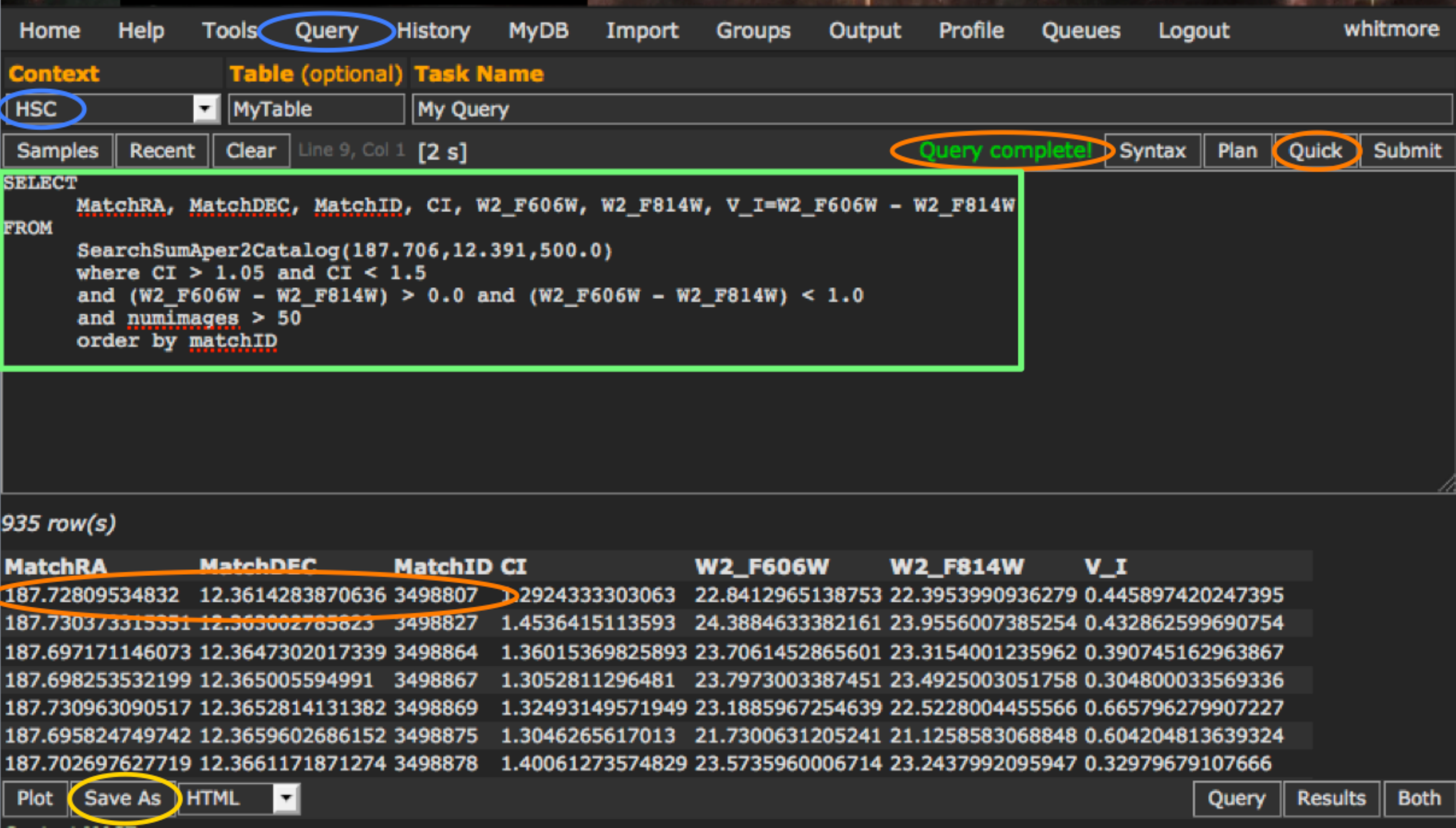}
\caption{Example of a HSC CasJobs screen from HSC Use Case~\#2.
\label{fig-CasJobs}}
\end{figure}

Figure~\ref{fig-SMC} shows an example of how CasJobs can be used to make a color magnitude diagram including 385,675 ACS 
sources in the Small Magellanic Cloud in less than two minutes (from Use Case~\#2). Figure~\ref{fig-CasJobs} shows an example 
of the query screen for CasJobs, in this case for retrieving a sample of globular clusters in M87 (also from HSC Use Case~\#2).

\subsection{HSC Home Page (Summary and Detailed Search Forms)}

The HSC Home Page (\url{http://archive.stsci.edu/hst/hsc}) represents a more basic level of sophistication. This was the original 
access tool (e.g., for the Beta releases), and while it is useful for certain very detailed searches, it has been largely superseded by the 
Discovery Portal and HSC CasJobs. It does, however, provide straightforward programmatic access to the HSC with a Virtual 
Observatory-compatible cone search and other scriptable interfaces to the HSC.  It also can be used for larger searches than the 
portal since it allows the selection of objects using parameters other than position (e.g., magnitudes).

There are two forms-based interfaces to the HSC linked from the Home Page that follow the conventions of MAST.  These are the 
summary search form, which allows users to obtain mean magnitudes and other information with one row per match, and the detailed 
search form, which includes information about each detection that went into the match. The HSC FAQ is also located at this site, 
providing the next level of detail beyond this paper.

\begin{figure}[t]
\centering
\includegraphics[type=\plotext,ext=.\plotext,read=.\plotext,width=0.75\linewidth]{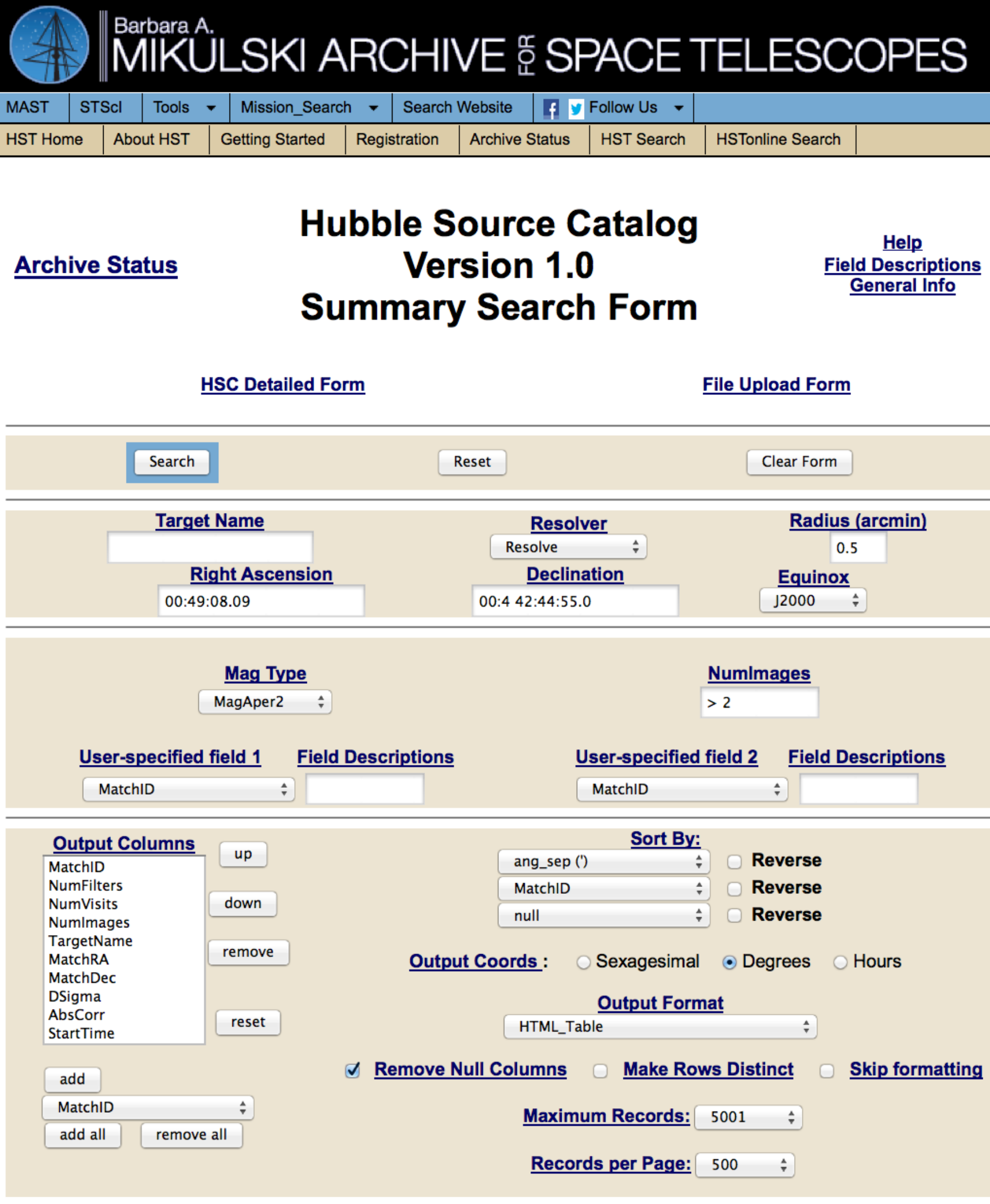}
\caption{Example of a search using the HSC Summary Search Form, accessible via the HSC Home Page.
\label{fig-form}}
\end{figure}

Figure~\ref{fig-form} shows an example of how the HSC Home Page can be used to download data from Brown et~al.\ (2009) 
observations of the outer disk of M31 (from HSC Use Case~\#1).

 \subsection{HLA Interactive Display (Image Browsing, Source Checking)}

The final way to access the HSC is via the Interactive Display within the HLA (\url{http://hla.stsci.edu}).  An ``advanced HSC 
controls'' feature allows the user to set a minimum value for \texttt{NumImages} in order to filter out cosmic rays and other artifacts. 
The HSC summary catalog information for a specific object can be displayed by clicking on the sources in the display. Examples of the 
Interactive Display are shown in Figures~\ref{fig-bad-WFPC2} and \ref{fig-nonuniform}. The Interactive Display can also be accessed 
through the Discovery Portal.
\newpage

\section{Example Parameter Files for Generation of HLA Source Lists}\label{sec:appendix_C}

\begin{verbatim}
#========================
# Sample ACS/WFC sextractor configuration file with specific values 
# for filter hst_10188_10_acs_wfc_f435w.

# Default configuration file for SExtractor 2.5.0
# EB 2010-12-20
# Adjusted with suitable defaults for one-orbit depth WFC3-IR images
#=======================
 
 #------------ Catalog --------------
CATALOG_NAME     hst_10188_10_acs_wfc_f435w_sex.cat
                                # name of the output catalog
CATALOG_TYPE     ASCII_HEAD     # NONE,ASCII,ASCII_HEAD, ASCII_SKYCAT,
                                # ASCII_VOTABLE, FITS_1.0 or FITS_LDAC
PARAMETERS_NAME  snpipe.sexparam  # name of the file containing catalog contents
 
#----------- Extraction ------------
DETECT_TYPE      CCD            # CCD (linear) or PHOTO (with gamma correction)
DETECT_MINAREA   5              # minimum number of pixels above threshold
THRESH_TYPE      RELATIVE       # threshold type: RELATIVE (in sigmas)
                                # or ABSOLUTE (in ADUs)
DETECT_THRESH    1.4            # <sigmas> or <threshold>,<ZP> in mag.arcsec-2
ANALYSIS_THRESH  4              # <sigmas> or <threshold>,<ZP> in mag.arcsec-2
 
FILTER           Y              # apply filter for detection (Y or N)?
FILTER_NAME      gauss_2.0_5x5.conv   # name of the file containing the filter
FILTER_THRESH                   # Threshold[s] for retina filtering
 
DEBLEND_NTHRESH  32             # Number of deblending sub-thresholds
DEBLEND_MINCONT  0.005          # Minimum contrast parameter for deblending
 
CLEAN            Y              # Clean spurious detections? (Y or N)?
CLEAN_PARAM      1.0            # Cleaning efficiency
 
MASK_TYPE        CORRECT        # type of detection MASKing: can be one of
                                # NONE, BLANK or CORRECT
 
#-------------WEIGHTing -----------
WEIGHT_TYPE      MAP_RMS        # type of WEIGHTing: NONE, BACKGROUND,
                                # MAP_RMS, MAP_VAR or MAP_WEIGHT
WEIGHT_IMAGE hst_10188_10_acs_wfc_iv2.fits,hst_10188_10_acs_wfc_f435w_rms.fits
                                # weight-map filename
WEIGHT_GAIN      N              # modulate gain (E/ADU) with weights? (Y/N)
WEIGHT_THRESH                   # weight threshold[s] for bad pixels

#------------ FLAGging -------------
FLAG_IMAGE       hst_10188_10_acs_wfc_f435w_msk.fits
                                # filename for an input FLAG-image
FLAG_TYPE        OR             # flag pixel combination: OR, AND, MIN, MAX
                                # or MOST

#---------- Photometry -------------
PHOT_APERTURES   2.0, 6.0       # MAG_APER aperture diameter(s) in pixels
PHOT_AUTOPARAMS  2.5, 3.5       # MAG_AUTO parameters: <Kron_fact>,<min_radius>
PHOT_PETROPARAMS 2.0, 3.5       # MAG_PETRO parameters: <Petrosian_fact>,
                                # <min_radius>
PHOT_AUTOAPERS   3.0,3.0        # <estimation>,<measurement> minimum apertures
                                # for MAG_AUTO and MAG_PETRO
PHOT_FLUXFRAC    0.5            # flux fraction[s] used for FLUX_RADIUS
 
SATUR_LEVEL      60000.0        # level (in ADUs) at which arises saturation
 
MAG_ZEROPOINT    25.6838624451  # magnitude zero-point
MAG_GAMMA        4.0            # gamma of emulsion (for photographic scans)
GAIN             2192.0         # detector gain in e-/ADU
PIXEL_SCALE      0.05           # size of pixel in arcsec (0=use FITS WCS info)
 
#-------- Star/Galaxy Separation ---------
SEEING_FWHM      0.076           # stellar FWHM in arcsec
STARNNW_NAME     hla_wfc3.nnw    # Neural-Network_Weight table filename
 
#---------- Background ------------
BACK_TYPE        AUTO           # AUTO or MANUAL
BACK_VALUE       0.0            # Default background value in MANUAL mode
BACK_SIZE        32            # Background mesh: <size> or <width>,<height>
BACK_FILTERSIZE  3              # Background filter: <size> or <width>,<height>
 
BACKPHOTO_TYPE   LOCAL          # can be GLOBAL or LOCAL
BACKPHOTO_THICK  32            # thickness of the background LOCAL annulus
BACK_FILTTHRESH  0.0            # Threshold above which the background-
                                # map filter operates
 
#---------Check Image ----------
CHECKIMAGE_TYPE BACKGROUND,APERTURES,SEGMENTATION,OBJECTS
                                # can be NONE, BACKGROUND, BACKGROUND_RMS,
                                # MINIBACKGROUND, MINIBACK_RMS, -BACKGROUND,
                                # FILTERED, OBJECTS, -OBJECTS, SEGMENTATION,
                                # or APERTURES
CHECKIMAGE_NAME hst_10188_10_acs_wfc_f435w_BGD_sex.fits,hst_10188_10_acs_wfc_f435w_APR_sex.fits,
hst_10188_10_acs_wfc_f435w_SGM_sex.fits,hst_10188_10_acs_wfc_f435w_OBJ_sex.fits
                                # Filename for the check-image
 
#----- Memory (change with caution!) ------
MEMORY_OBJSTACK  60000           # number of objects in stack
MEMORY_PIXSTACK  10000000         # number of pixels in stack
MEMORY_BUFSIZE   1024           # number of lines in buffer
 
#------- ASSOCiation --------
ASSOC_NAME       sky.list       # name of the ASCII file to ASSOCiate
ASSOC_DATA       2,3,4          # columns of the data to replicate (0=all)
ASSOC_PARAMS     2,3,4          # columns of xpos,ypos[,mag]
ASSOC_RADIUS     2.0            # cross-matching radius (pixels)
ASSOC_TYPE       MAG_SUM        # ASSOCiation method: FIRST, NEAREST, MEAN,
                                # MAG_MEAN, SUM, MAG_SUM, MIN or MAX
ASSOCSELEC_TYPE  MATCHED        # ASSOC selection type: ALL, MATCHED or -MATCHED

#--------- Miscellaneous ----------
VERBOSE_TYPE     NORMAL         # can be QUIET, NORMAL or FULL
WRITE_XML        N              # Write XML file (Y/N)?
XML_NAME         sex.xml        # Filename for XML output
XSL_URL          file:///usr/local/share/sextractor/sextractor.xsl
                                # Filename for XSL style-sheet
NTHREADS         1              # 1 single thread

FITS_UNSIGNED    N              # Treat FITS integer values as unsigned (Y/N)?
INTERP_MAXXLAG   16             # Max. lag along X for 0-weight interpolation
INTERP_MAXYLAG   16             # Max. lag along Y for 0-weight interpolation
INTERP_TYPE      ALL            # Interpolation type: NONE, VAR_ONLY or ALL

#-----------------------------

\end{verbatim}

\section{Access to Information}\label{sec:appendix_D}

The four primary ways to access the HSC are:

\begin{itemize}
\item MAST Discovery Portal---\url{http://mast.stsci.edu}

\item HSC CasJobs---\url{http://mastweb.stsci.edu/hcasjobs}

\item HSC Home Page and Search Forms---\url{http://archive.stsci.edu/hst/hsc/}

\item HLA Interactive Display----\url{http://hla.stsci.edu}
\end{itemize}

Other sources of more detailed information include:

\begin{itemize}
\item HSC Use Cases \url{http://archive.stsci.edu/hst/hsc/help/HSC_faq.html#use_case}

\item HSC FAQ \url{http://archive.stsci.edu/hst/hsc/help/HSC_faq.html}

\item HLA FAQ \url{http://hla.stsci.edu/hla_faq.html}

\item Discovery Portal User's Guide
\url{http://mast.stsci.edu/portal/Mashup/Clients/Mast/data/html/MastHelp.html}

\item HSC CasJobs Guide \url{http://mastweb.stsci.edu/hcasjobs/guide.aspx}

\item Space Telescope Science Institute (STScI) Archive Help Desk:  archive@stsci.edu
\end{itemize}

\end{document}